\documentclass[preprint,12pt]{elsarticle}

\usepackage[margin=1in]{geometry}
\usepackage{cool}
\usepackage{hyperref}
\usepackage{multirow}
\usepackage{caption}
\usepackage{subcaption}
\usepackage{tikz}
\usepackage{mathtools}
\usepackage{nicematrix}
\usepackage{subfiles}
\usepackage{cool}
\usepackage{hyperref}
\usepackage{caption}
\usepackage{stmaryrd}
\usepackage{natbib}
\usepackage{booktabs}
\newcommand{\e}{\mathrm{e}}
\newcommand{\diff}[1]{\mathrm{d}#1}
\renewcommand{\Re}{\mathrm{Re}}
\renewcommand{\Im}{\mathrm{Im}}

\newcommand{\upi}{\mathrm{i}}

\newcommand{\w}{\mathrm{w}}
\newcommand{\p}{\mathrm{p}}
\newcommand{\q}{\mathrm{q}}
\newcommand{\vo}{\mathrm{v}}
\newcommand{\upd}{\mathrm{d}}

\makeatletter
\let\DOTSI\relax 
\newcommand*{\Xint}{%
  \DOTSI
  \mathop{%
    \mathpalette\@LetterOnInt{\times}%
  }%
  \mkern-\thinmuskip 
  \int
}
\newcommand*{\Uint}{%
  \DOTSI
  \mathop{%
    \mathpalette\@LetterOnInt{\smile}%
  }%
  \mkern-\thinmuskip 
  \int
}
\newcommand*{\@LetterOnInt}[2]{%
  \sbox0{$#1\int\m@th$}%
  \sbox2{$%
    \ifx#1\displaystyle
      \textstyle
    \else
      \scriptscriptstyle
    \fi
    #2%
  \m@th$}%
  \dimen@=.4\dimexpr\ht0+\dp0\relax
  \ifdim\dimexpr\ht2+\dp2\relax>\dimen@
    \sbox2{\resizebox*{!}{\dimen@}{\unhcopy2}}%
  \fi
  \dimen@=\wd0 %
  \ifdim\wd2>\dimen@
    \dimen@=\wd2 %
  \fi
  \rlap{\hbox to \dimen@{\hfil
    $#1\vcenter{\copy2}\m@th$%
  \hfil}}%
  \ifdim\dimen@>\wd0 %
    \kern.5\dimexpr\dimen@-\wd0\relax
  \fi
}
\makeatother

\newcommand{\ii}[1]{\mathrm{i}}

\usepackage[dvipsnames]{xcolor}
\newcommand{\change}[1]{#1}
\newcommand{\changerevtwo}[1]{#1}

\journal{Renewable energy}

\begin{document}

\begin{frontmatter} 

\title{Wave energy conversion by floating and submerged piezoelectric bimorph plates}


\author[inst3]{Zachary J. Wegert}
\author[inst2,inst5]{Ben Wilks}
\author[inst2]{Ngamta Thamwattana}
\author[inst3]{Vivien J. Challis}
\author[inst4]{Santanu Koley}
\author[inst2]{Michael H. Meylan}


\affiliation[inst3]{organization={School of Mathematical Sciences},
            addressline={Queensland University of Technology}, 
            city={Brisbane}, 
            state={Queensland},
            postcode={4001},
            country={Australia}}

\affiliation[inst2]{organization={School of Computer and Information Sciences},
            addressline={University of Newcastle}, 
            city={Callaghan}, 
            state={New South Wales},
            postcode={2308},
            country={Australia}}

\affiliation[inst5]{organization={School of Mathematical Sciences},
            addressline={Adelaide University}, 
            city={Mawson Lakes}, 
            state={South Australia},
            postcode={5095},
            country={Australia}}

\affiliation[inst4]{organization={Department of Mathematics},
            addressline={Birla Institute of Technology and Science-Pilani, Hyderabad Campus}, 
            city={Hyderabad},
            state={Telangana},
            postcode={500078},
            country={India}}

\begin{abstract}
Gaining insight into the interaction between flexible piezoelectric structures and ocean waves can inform the development of compact, high-efficiency wave-energy converters that harvest renewable energy from the marine environment. In this paper, the problem of wave energy absorption by floating and submerged piezoelectric plates is investigated. The equations of motion for a plate consisting of two piezoelectric layers separated by an elastic substrate are derived in dimensional form from the full piezoelectric constitutive laws. \change{A solution using a modal expansion method is proposed, in which the component radiation and diffraction problems are reduced to hypersingular integral equations and solved numerically using a constant panel method. The method is general and can solve the equations of motion for submerged rigid, flexible elastic or flexible piezoelectric plates.} Extensive numerical results\change{ for the energy absorption and efficiency } are given for a range of parameters, including different piezoelectric materials: polyvinylidene fluoride (PVDF) and lead zirconate titanate (PZT-5H). Importantly, greater energy absorption is obtained for submerged plates when compared to plates floating on the surface. Furthermore, clamped boundary conditions give slightly larger energy absorption compared to the simply supported case. Our open-source code is provided at \url{https://github.com/zjwegert/SemiAnalyticWECs.jl}.
\end{abstract}



\end{frontmatter}

\section{Introduction}

The efficient and cost-effective extraction of ocean wave energy is the subject of extensive current research \cite{Falcao,Falnes}. Numerous designs for wave energy converters (WECs) have been built or proposed \cite{renzi2021niche}, and one type that has been the subject of recent investigation is piezoelectric WECs (PWECs). Several recent works have given an overview of technologies that use piezoelectric materials to convert the mechanical motion of ocean waves into electrical energy and discussed different types of structures, piezoelectric materials,  their operating principles, advantages, and limitations \cite{Review_Kiran_etal2020,Review_Kargar_Hao_2022,Review_Liu_etal2023,Review_Li_etal2024,Review_Wang_etal2024,Review_Xiao_2025}. In particular, 
\citet{Review_Kargar_Hao_2022} provide a comprehensive compilation of 85 piezoelectric energy harvester designs tailored for oceanic applications, which are categorised based on their configurations (e.g., cantilever beam, diaphragm, stacked, and cymbal), materials, coupling modes, operational locations, and power output ranges. In terms of mathematical modelling, the first papers to investigate submerged elastic plates were theoretical extensions of the work on floating elastic plates, and had no clear application \cite{ISOPEjnl1995,meylan2009water, williams_meylan12}. Building on this work, \citet{renzi2016hydroelectromechanical} was one of the first authors to analytically couple the hydroelectromechanical behaviour of such systems. The paper presents a fully coupled theoretical model that integrates fluid dynamics, structural mechanics, and piezoelectric effects to describe how a flexible plate with embedded piezoelectric layers can convert ocean wave energy into electricity.

Based on Renzi’s work \cite{renzi2016hydroelectromechanical}, PWECs have been extensively studied with particular attention to the placement of piezoelectric plates (floating versus submerged). Here, we first discuss recent studies of submerged piezoelectric plates.  \citet{Zheng_etal2021,Zheng_etal2021_proc} analysed a WEC comprised of a submerged flexible plate with piezoelectric layers, mounted in front of a floating breakwater, using a coupled hydro-elastic and electromechanical model based on linear potential flow and eigenfunction matching. Based on their multi-parameter study, \citet{Zheng_etal2021} concluded that deeper submergence of the plate causes the main peaks in the frequency response of the absorption efficiency to become lower in magnitude and narrower in bandwidth. Additionally, \citet{Zheng_etal2021_proc} found that increasing the width of the PWEC can lead to more peaks in the frequency response of the wave power absorption efficiency. \citet{Vipin_Koley_2022} considered a WEC  comprising a flexible submerged plate with piezoelectric layers attached on both faces, mounted over a seabed with variable (undulated) topography. Using potential-flow theory combined with Euler-Bernoulli plate theory and  a boundary‐element method to solve the boundary‐value problem, they showed that seabed undulation and plate boundary conditions significantly influence the hydrodynamic excitation and therefore the predicted energy harvesting potential. The seabed topography and structural tuning in submerged piezoelectric WEC design is also topic of consideration in other recent work \cite{Vipin_etal_2022,Trivedi_Koley_2023b,Vipin_Koley_2024}. 

Recently, \citet{koley2025dualpiezo} extended the earlier work of  \citet{Vipin_Koley_2022} by considering two horizontal piezoelectric plates over an undulated seabed. The results demonstrate that dual piezoelectric plates are capable of generating more power than the single piezoelectric plate system. \citet{Trivedi_Koley_2023} focus on optimising the performance of PWECs by employing the Taguchi method (a multi-factor optimisation process) to identify the most effective design parameters for the devices. The Taguchi method has also been used to study a submerged piezoelectric plate integrated with a breakwater over an undulated seabed \cite{Vipin_Koley_2024} and a hybrid WEC combining a flexible piezoelectric plate and an oscillating water column  device mounted over an undulated seabed \cite{Trivedi_Koley_2023b}. \citet{Shoele2023} investigated a flexible piezoelectric plate submerged near the free surface subjected to both gravity waves and ocean current using an inviscid‐fluid model to capture the hydrodynamic coupling between the plate, incident waves, and current. The study indicates that optimal performance occurs when the plate is submerged to a depth less than half its length. Using an artificial neural network  model to optimise the design  of a submerged PWEC, \citet{Vipin_Koley_ANN2023} found that maximum power can be extracted when the ratio of the plate half-length to the ocean depth is approximately $2.7$--$2.9$ and the submergence depth is approximately 1.2\% of the ocean depth. We note that instead of a rectangular plate, other geometries of submerged piezoelectric plate WECs have also been considered, such as a circular cylinder \citep{Shen_etal2024} and a U-shaped geometry \citep{Ushape}. \change{It should be noted that although PWECs are a promising method for power generation, currently proposed PWECs lack broadband energy capture and are not highly efficient across large frequency ranges. It may be possible to resolve this by having multiple devices that are tuned to different frequency ranges as investigated by \citet{Westcott_Bennetts_Sergiienko_Cazzolato_2024} for buoy WECs. In the context of PWECs, tuning the frequency range could be achieved by adjusting parameters including the submergence depth, surface conductivity, and poling angle. We therefore consider varying these quantities in our numerical results.}
 
Although much research has focused on submerged plates due to the potentially higher power generation, there are also a number of recent works that focus on floating WECs. 
\citet{Meylan_Challis_Thamwattana_Wegert_Wilks_2025} investigated the wave scattering, motion and energy absorption of floating elastic plates with
an imaginary component to their bending rigidity. The imaginary part was included because energy can be extracted from bending by leveraging the piezoelectric effect. More recently, the hydrodynamic component of that work \cite{Meylan_Challis_Thamwattana_Wegert_Wilks_2025} was extended to three-dimensional water waves, in particular considering rectangular anisotropic plates on the fluid surface, however the application to PWECs was not considered directly \citep{wilks2025waterwavescatteringsurfacemounted}. For the study of floating plates that incorporate the piezoelectric effect, \citet{Li_etal2025} investigated the hydroelastic behaviour of a piezoelectric plate placed on the wave surface using viscous flow theory and coupled it with a finite element approach for the elastic deformation of the plate and the calculation of the static electric field. They chose surface placement to increase the static pressure differential between the top and bottom of the plate and amplify the wave‑induced loads. Although the above papers consider rectangular plates, there are other configurations of floating PWECs that are designed to improve efficiency in energy harvesting (see, for example, the work of   \citet{Chen_etal2021,Chen_etal2021b,Cai_etal2021,Cshape}). We also note that \citet{zheng2020water} and \citet{michele2022wave} have considered wave energy conversion by circular flexible plates that are either submerged or surface mounted, respectively, however they considered power takeoff mechanisms unrelated to piezoelectricity. Piezoelectric effects have not yet been considered for circular plates. 

Renzi's early hyrdoelectromechanical analysis \cite{renzi2016hydroelectromechanical} presented non-dimensionalised governing equations for a submerged bimorph PWEC comprising two oppositely-poled piezoelectric layers separated by an elastic substrate. Our derivation in this paper follows the key steps of \citet{renzi2016hydroelectromechanical}, however we derive the governing equations in dimensional form and begin with the full, three-dimensional piezoelectric constitutive laws. We provide this full derivation relevant to both surface and submerged PWECs and fill in additional details. 
The form of our derivation is particularly advantageous because it can readily be extended to three dimensions, and in this paper we incorporate an arbitrary rotation of the poling angle for the two oppositely-poled piezoelectric layers in the PWEC. We consider two types of piezoelectric materials, namely polyvinylidene fluoride (PVDF) and lead zirconate titanate (PZT-5H). Both PVDF and PZT-5H are commonly used in PWEC applications \cite{Review_Kargar_Hao_2022}. As discussed by \citet{Review_Kargar_Hao_2022}, PZT-type materials can produce more electricity due to their large piezoelectric coupling, while PVDF possesses a better load-bearing capacity.

\change{The objectives of this paper are to i) present a detailed derivation of the governing equations for PWECs, including an extension to an arbitrary poling angle; ii) develop a novel and efficient solution method based on a hypersingular boundary integral equation; and iii) evaluate and compare the power conversion performance of PVDF and PZT-5H as well as different parameter regimes.} The outline of the paper is as follows. In \S\ref{theory} we outline the mathematical theory of a submerged piezoelectric bimorph. In \S\ref{solution_sec} we present a novel solution method using a matrix presentation of the fluid operator, mimicking the standard method in general hydroelasticity. In \S\ref{numerics_sec} we give a detailed numerical study, focusing on realistic piezoelectric materials. We end with a brief conclusion in \S\ref{conclusions_sec}, followed by three appendices that give, in turn, the derivations of the material constants, the complex analytic method used to calculate the Green's function numerically, and a validating calculation that can serve as a benchmark for submerged elastic plates.  

\section{Mathematical theory}
\label{theory}
\subsection{Equations of motion for a piezoelectric bimorph}
\begin{figure}[htbp]
    \centering
    \def\svgwidth{0.7\textwidth}
    \large
    \input{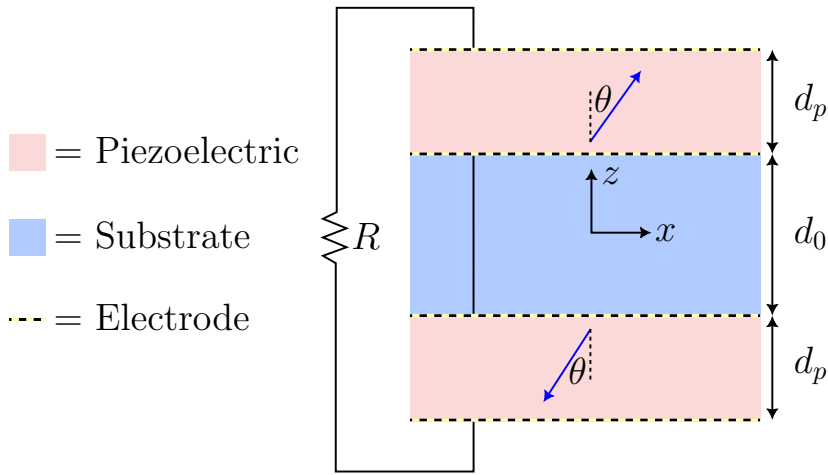}
    \caption{Visualisation of a piezoelectric bimorph circuited in series comprising two oppositely poled piezoelectric layers with arbitrary poling direction $\theta$, resistance $R$, and thicknesses $d_p$ and $d_0$ for the piezoelectric layers (pink) and elastic substrate (blue), respectively. The black dashed lines represent the electrodes on the top and bottom of each piezoelectric plate.}
    \label{fig:arb-poling-angle-series}
\end{figure}

\noindent In the following, we derive the equations of motion for a piezoelectric bimorph arranged in series as shown in Figure \ref{fig:arb-poling-angle-series}. The configuration is similar to \citet{renzi2016hydroelectromechanical}, with extension to include an arbitrary rotation of the poling angle for the piezoelectric layers. The derivation is also based on the approach of \citet{renzi2016hydroelectromechanical}, however, we instead give the derivation in dimensional form. We also provide more details, including beginning the derivation with the full material tensors that describe a three-dimensional piezoelectric material.

\subsubsection{Constitutive laws}
\noindent As indicated in Figure~\ref{fig:arb-poling-angle-series}, the two plates in the piezoelectric bimorph have opposite poling directions that relate to the alignment of electric dipoles within each piezoelectric layer. A poling angle of $\theta = 0^\circ$ relates to the vertical poling considered by \citet{renzi2016hydroelectromechanical}, and to consider arbitrary $\theta$ we apply an appropriate coordinate transformation to the base linear piezoelectric material constants for each of the piezoelectric layers. 

We introduce the following tensors describing the behaviour of the base piezoelectric material: a compliance tensor $\bar{S}^E_{ijkl}$, a piezoelectric coupling tensor $\bar{d}_{ijk}$, and a dielectric tensor $\bar{\kappa}^\sigma_{ij}$. The superscripts $E$ and $\sigma$ that appear on the compliance tensor and dielectric tensor denote that these tensors are specified for constant electric field and stress, respectively. In Voigt notation, the three piezoelectric tensors for an orthotropic piezoelectric material poled in the $z$ direction can be written as \citep{YangIntroPZ}
\begin{subequations}
\begin{align}
    \bar{\boldsymbol{S}}^E &= \begin{pmatrix}
        \bar{S}^E_{11} & \bar{S}^E_{12} & \bar{S}^E_{13} & 0 & 0 & 0\\
        \bar{S}^E_{12} & \bar{S}^E_{22} & \bar{S}^E_{23} & 0 & 0 & 0\\
        \bar{S}^E_{13} & \bar{S}^E_{23} & \bar{S}^E_{33} & 0 & 0 & 0\\
        0 & 0 & 0 & \bar{S}^E_{44} & 0 & 0\\
        0 & 0 & 0 & 0 & \bar{S}^E_{55} & 0\\
        0 & 0 & 0 & 0 & 0 & \bar{S}^E_{66}
    \end{pmatrix},\\
    \bar{\boldsymbol{d}} &= \begin{pmatrix}
        0&0&0&0&\bar{d}_{15}&0\\
        0&0&0&\bar{d}_{24}&0&0\\
        \bar{d}_{31}&\bar{d}_{32}&\bar{d}_{33}&0&0&0
    \end{pmatrix},\\
    \bar{\boldsymbol{\kappa}}^\sigma &=\begin{pmatrix}
        \bar{\kappa}^\sigma_{11}&0&0\\
        0&\bar{\kappa}^\sigma_{22}&0\\
        0&0&\bar{\kappa}^\sigma_{33}
    \end{pmatrix},
\end{align}
\end{subequations}
where the specific values of the nonzero components relate to the piezoelectric material being used.

When the piezoelectric material is rotated by an angle of $\theta$ in the $xz$-plane, the piezoelectric material tensors transform as 
\begin{subequations}
\begin{align}
    &S^E_{ijkl}(\theta)=R_{im}(\theta)R_{jl}(\theta)R_{ko}(\theta)R_{lp}(\theta)\bar{S}^E_{mlop},\\
    &d_{ijk}(\theta)=R_{im}(\theta)R_{jl}(\theta)R_{ko}(\theta)\bar{d}_{mlo},\\
    &\kappa^\sigma_{ij}(\theta)=R_{im}(\theta)R_{jl}(\theta)\bar{\kappa}_{ml}^\sigma,
\end{align}
\end{subequations}
where Einstein's summation convention is followed and $R_{ij}(\theta)$ are the components of a standard clockwise coordinate rotation given by
\begin{equation}
    \boldsymbol{R}(\theta)=\begin{pmatrix}
        \cos(\theta)&0&\sin(\theta)\\
        0&1&0\\
        -\sin(\theta)&0&\cos(\theta)
    \end{pmatrix}.
\end{equation}
This yields the following non-zero components for the piezoelectric material tensors:
\begin{subequations}
\begin{align}
    \boldsymbol{S}^E(\theta) &= \begin{pmatrix}
        {S}^E_{11}(\theta) & {S}^E_{12}(\theta) & {S}^E_{13}(\theta) & 0 & {S}^E_{15}(\theta) & 0\\
        {S}^E_{12}(\theta) & {S}^E_{22}(\theta) & {S}^E_{23}(\theta) & 0 & {S}^E_{25}(\theta) & 0\\
        {S}^E_{13}(\theta) & {S}^E_{23}(\theta) & {S}^E_{33}(\theta) & 0 & {S}^E_{35}(\theta) & 0\\
        0 & 0 & 0 & {S}^E_{44}(\theta) & 0 & {S}^E_{46}(\theta)\\
        {S}^E_{15}(\theta) & {S}^E_{25}(\theta) & {S}^E_{35}(\theta) & 0 & {S}^E_{55}(\theta) & 0\\
        0 & 0 & 0 & {S}^E_{46}(\theta) & 0 & {S}^E_{66}(\theta)
    \end{pmatrix},\\
    \boldsymbol{d}(\theta) &= \begin{pmatrix}
        {d}_{11}(\theta)&{d}_{12}(\theta)&{d}_{13}(\theta)&0&{d}_{15}(\theta)&0\\
        0&0&0&{d}_{24}(\theta)&0&{d}_{26}(\theta)\\
        {d}_{31}(\theta)&{d}_{32}(\theta)&{d}_{33}(\theta)&0&{d}_{35}(\theta)&0
    \end{pmatrix},\\
    \boldsymbol{\kappa}^\sigma(\theta) &=\begin{pmatrix}
        {\kappa}^\sigma_{11}(\theta)&0&{\kappa}^\sigma_{13}(\theta)\\
        0&{\kappa}^\sigma_{22}(\theta)&0\\
        {\kappa}^\sigma_{13}(\theta)&0&{\kappa}^\sigma_{33}(\theta)
    \end{pmatrix}.
\end{align}
\end{subequations}

We now can specify the constitutive law for the top piezoelectric layer shown in Figure~\ref{fig:arb-poling-angle-series} in Voigt notation as \citep{YangIntroPZ}
\begin{equation}\label{eqn: strain-electric disp}
    \begin{pNiceArray}{c}[cell-space-limits=3pt]
  \boldsymbol{\varepsilon}\\\boldsymbol{D}
\end{pNiceArray}=\begin{pNiceArray}{c|c}[cell-space-limits=3pt]
  \boldsymbol{S}^E(\theta) & \boldsymbol{d}(\theta)^\intercal \\
  \hline
  \boldsymbol{d}(\theta) & \boldsymbol{\kappa}^\sigma(\theta)
\end{pNiceArray}\begin{pNiceArray}{c}[cell-space-limits=3pt]
  \boldsymbol{\sigma}\\\boldsymbol{E}
\end{pNiceArray},
\end{equation}
where
\begin{equation}
    \boldsymbol{\varepsilon}=(\varepsilon_{11},\varepsilon_{22},\varepsilon_{33},2\varepsilon_{23},2\varepsilon_{13},2\varepsilon_{12})
\end{equation}
and
\begin{equation}
    \boldsymbol{\sigma}=(\sigma_{11},\sigma_{22},\sigma_{33},\sigma_{23},\sigma_{13},\sigma_{12})
\end{equation}
are the Voigt notation forms of the strain tensor  $\varepsilon_{ij}$ and the stress tensor $\sigma_{ij}$, respectively. In addition, $\boldsymbol{E}$ is the electric field vector and $\boldsymbol{D}$ is the electric displacement vector. 

For the elastic substrate, we assume that the material is linear with a compliance tensor $S^0_{ijkl}$ that has the same symmetries as $\bar{S}^E_{ijkl}$. The constitutive law for the elastic layer is given by 
\begin{equation}\label{eqn: strain-stress}
    \boldsymbol{\varepsilon}=\boldsymbol{S}^0\boldsymbol{\sigma},
\end{equation}
where $\boldsymbol{S}^0$ is the Voigt matrix form for $S_{ijkl}^0$. 

We make several assumptions to simplify these three-dimensional constitutive laws. Following \citet{alma991005216849704001}, if we assume that the plate is thin in the $z$-direction and that the electrodes cover the faces perpendicular to the $z$-direction, we have plane stress conditions 
\begin{equation}
    \sigma_{33}=\sigma_{23}=\sigma_{13}=0,
\end{equation}
and electrical conditions
\begin{equation} \label{eqn:zeroE1E2}
    E_1=E_2=0.
\end{equation}
As a result, Equations \eqref{eqn: strain-electric disp} and \eqref{eqn: strain-stress} can be written as 
\begin{equation}\label{eqn: strain-electric disp simplif}
    \begin{pmatrix}
        \varepsilon_{11}\\\varepsilon_{22}\\2\varepsilon_{12}\\D_3
    \end{pmatrix}=
    \begin{pmatrix}
        S^E_{11}(\theta)&S^E_{12}(\theta)&0&d_{31}(\theta)\\
        S^E_{12}(\theta)&S^E_{22}(\theta)&0&d_{32}(\theta)\\
        0&0&S^E_{66}(\theta)&0\\
        d_{31}(\theta)&d_{32}(\theta)&0&\kappa^\sigma_{33}(\theta)
    \end{pmatrix}
    \begin{pmatrix}
        \sigma_{11}\\\sigma_{22}\\\sigma_{12}\\E_3
    \end{pmatrix},
\end{equation}
and
\begin{equation}\label{eqn: strain-stress simplif}
    \begin{pmatrix}
        \varepsilon_{11}\\\varepsilon_{22}\\2\varepsilon_{12}
    \end{pmatrix}=
    \begin{pmatrix}
        S^E_{11}&S^E_{12}&0\\
        S^E_{12}&S^E_{22}&0\\
        0&0&S^E_{66}\\
    \end{pmatrix}
    \begin{pmatrix}
        \sigma_{11}\\\sigma_{22}\\\sigma_{12}
    \end{pmatrix},
\end{equation}
respectively. Rearranging Equation \eqref{eqn: strain-electric disp simplif} to be in stress-electric displacement form and Equation \eqref{eqn: strain-stress simplif} to be in stress form yields
\begin{equation}\label{eqn: strain-electric disp simplif2}
    \begin{pmatrix}
        \sigma_{11}\\\sigma_{22}\\\sigma_{12}\\D_3
    \end{pmatrix}=
    \begin{pmatrix}
        S^E_{11}(\theta)&S^E_{12}(\theta)&0&0\\
        S^E_{12}(\theta)&S^E_{22}(\theta)&0&0\\
        0&0&S^E_{66}(\theta)&0\\
        -d_{31}(\theta)&-d_{32}(\theta)&0&1
    \end{pmatrix}^{-1}
    \begin{pmatrix}
        1&0&0&-d_{31}(\theta)\\
        0&1&0&-d_{32}(\theta)\\
        0&0&1&0\\
        0&0&0&\kappa^\sigma_{33}(\theta)
    \end{pmatrix}
    \begin{pmatrix}
        \varepsilon_{11}\\\varepsilon_{22}\\2\varepsilon_{12}\\E_3
    \end{pmatrix},
\end{equation}
and
\begin{equation}\label{eqn: strain-stress simplif2}
    \begin{pmatrix}
        \sigma_{11}\\\sigma_{22}\\\sigma_{12}
    \end{pmatrix}=
    \begin{pmatrix}
        S^E_{11}&S^E_{12}&0\\
        S^E_{12}&S^E_{22}&0\\
        0&0&S^E_{66}\\
    \end{pmatrix}^{-1}
    \begin{pmatrix}
        \varepsilon_{11}\\\varepsilon_{22}\\2\varepsilon_{12}
    \end{pmatrix},
\end{equation}
respectively. Finally, we assume that both the forcing due to the fluid and the bending of the plate will be invariant in $y$. We therefore can assume that we have plane strain in $y$, as in \cite{Reddy_2006}, yielding the strain conditions
\begin{equation}
    \varepsilon_{22}=\varepsilon_{12}=0. 
\end{equation}
The constitutive law for the top thin piezoelectric plate becomes
\begin{subequations}
\begin{align}
    \sigma_{11}&=\hat{C}_{11}^E(\theta)\varepsilon_{11}-\hat{e}_{31}(\theta)E_3,\label{eqn: topplate sigma}\\
    D_3&=\hat{e}_{31}(\theta)\varepsilon_{11}+\hat{\kappa}^\varepsilon_{33}(\theta)E_3,\label{eqn: topplate disp}
\end{align}
\end{subequations}
and the constitutive law for the elastic substrate is
\begin{equation} \label{eqn: elastic sigma}
    \sigma_{11}=\hat{C}^0_{11}\varepsilon_{11},
\end{equation}
where
\begin{subequations}
\begin{align}
    \hat{C}_{11}^E(\theta)&=\frac{S^E_{22}(\theta)}{S_{11}^E(\theta)S_{22}^E(\theta)-S_{12}^E(\theta)^2},\\
    \hat{e}_{31}(\theta)&=\frac{d_{32}(\theta)S^E_{12}(\theta)-d_{31}S_{22}^E(\theta)}{S_{11}^E(\theta)S_{22}^E(\theta)-S_{12}^E(\theta)^2},\\
    \hat{\kappa}^\varepsilon_{33}(\theta)&=\frac{d_{32}(\theta)^2 S^E_{11}(\theta) - 2 d_{31}(\theta) d_{32}(\theta) S^E_{12}(\theta) + d_{31}(\theta)^2 S^E_{22}(\theta) - \kappa^\sigma_{33}(\theta)[S^E_{11}(\theta) S^E_{22}(\theta)-S^E_{12}(\theta)^2]}{S_{11}^E(\theta)S_{22}^E(\theta)-S_{12}^E(\theta)^2},\\
    \hat{C}_{11}^0&=\frac{S^0_{22}}{S_{11}^0S_{22}^0-(S_{12}^0)^2}.
\end{align}
\end{subequations}
In the above, we use a hat to denote coefficients that have been reduced from three dimensions using plane stress conditions. The expressions for these in terms of components of the piezoelectric material tensors in the original reference frame are given in ~\ref{appendix1}.

Note that for the series bimorph in Figure \ref{fig:arb-poling-angle-series}, the poling direction of the bottom plate is opposite to the top plate. This corresponds to a phase shift of $180^\circ$ in the poling angle for the piezoelectric material constants of the bottom plate. As a result, one can show that on the bottom plate, Equations \eqref{eqn: topplate sigma} and \eqref{eqn: topplate disp} become
\begin{subequations}
\begin{align}
    \sigma_{11}&=\hat{C}_{11}^E(\theta)\varepsilon_{11}+\hat{e}_{31}(\theta)E_3,\label{eqn: bottomplate sigma}\\
    D_3&=-\hat{e}_{31}(\theta)\varepsilon_{11}+\hat{\kappa}^\varepsilon_{33}(\theta)E_3.\label{eqn: bottomplate disp}
\end{align}
\end{subequations}

\subsubsection{Circuiting}
\noindent Before we consider the equations for the bending of the plate and the resulting charge, we must first derive the equations that relate the voltage to the electric field. Connecting piezoelectric devices in series or in parallel controls whether voltage or charge is accumulated, respectively \citep{Liu_Zhong_Lee_Lee_Lin_2018,Wang_Du_Xu_Cross_1999,EnergyHarvesters_2011}. The piezoelectric bimorph shown in Figure~\ref{fig:arb-poling-angle-series} is connected in series through an elastic substrate. The electrodes are assumed to cover the entire top and bottom of each piezoelectric layer. With the layers connected in series, the voltage across the resistor is the sum of the voltages across the plates. Furthermore, the way the two piezoelectric layers are oppositely poled results in the same voltage across the top and bottom piezoelectric layers. That is,
\begin{equation}
    V(x,t)=2V_p(x,t),
\end{equation}
where $V(x,t)$ is the voltage across the resistor and $V_p(x,t)$ is the voltage across the top piezoelectric plate. Furthermore, $V_p(x,t)$ is given by
\begin{equation}\label{eqn: general voltage}
    V_p(x,t)=-\int_\gamma\boldsymbol{E}(x,t)\cdot\diff{\boldsymbol{r}},
\end{equation}
where $\boldsymbol{E}(x,t)$ is the electric field and $\boldsymbol{r}$ parameterises the curve $\gamma$ that connects the bottom and top electrodes on either side of the piezoelectric layer \cite{renzi2016hydroelectromechanical}. The electric field does not vary with $z$ due to the assumption that the electrodes cover the top and bottom of the piezoelectric layers \citep{Erturk_Inman_2008}.  The two components $E_1$ and $E_2$ are zero (see Equation \eqref{eqn:zeroE1E2}), and the integral can readily be computed as a one-dimensional integral to give
\begin{equation}
    V_p(x,t)=-E_3(x,t)d_p,
\end{equation}
and therefore
\begin{equation}\label{eqn: voltage 1 bimorph}
    V(x,t)=-2E_3(x,t)d_p.
\end{equation}

\subsubsection{Bending}
\noindent We can now derive the equation for plate bending in terms of displacement, voltage, and surface load. Assuming that the layers are perfectly bonded, applying a Kirchhoff thin plate assumption, and neglecting $x$ and $y$ axial displacements, the equation of motion of the plate is given by \cite{Reddy_2006}
\begin{equation}\label{eqn: motion}
    \partial_x^2{M_{11}}-I_b\partial^2_{t}{W}=-P,
\end{equation}
where $W(x,t)$ is the out-of-plane displacement, $P(x,t)$ is the surface load perpendicular to the plate, and $M_{11}$ is the bending moment given by
\begin{equation}\label{eqn:bending mom}
    M_{11}=\int_{-d_b/2}^{d_b/2}z\sigma_{11}\,\mathrm{d}z,
\end{equation}
where $d_b=d_0+2d_p$ is the total height of the plate (refer to Figure~\ref{fig:arb-poling-angle-series}). Furthermore, $I_b$ is the first mass moment of inertia given by
\begin{equation} \label{eqn: first mass moment}
    I_b=\int_{-d_b/2}^{d_b/2}\rho(z)\,\mathrm{d}z=d_0\rho_0+2d_p\rho_p,
\end{equation}
where $\rho_0$ and $\rho_p$ are the constant densities of the elastic and piezoelectric materials,  respectively. 

In the absence of in-plane displacements, $\varepsilon_{11}$ is related to the out-of-plane displacement via \cite{Reddy_2006}
\begin{align}\label{eqn:strain relationship}
    \varepsilon_{11}=-z\partial^2_{x}{W}.
\end{align}
The constitutive relations then give the stress for the top piezoelectric plate and elastic substrate as
\begin{equation}\label{sigma_11}
\sigma_{11} = -z\hat{C}_{11}^E(\theta)\partial^2_{x}{W} - \hat{e}_{31}(\theta) E_3,
\end{equation}
and
\begin{equation}\label{eqn: stress elast}
    \sigma_{11} = -z\hat{C}^0_{11}\partial^2_{x}{W},
\end{equation}
respectively, which have been obtained simply by putting Equation \eqref{eqn:strain relationship} into the two constitutive relations Equation \eqref{eqn: topplate sigma} and Equation \eqref{eqn: elastic sigma}. Using Equation \eqref{eqn: voltage 1 bimorph} that relates the electric field component $E_3$ to the voltage across the resistor, Equation \eqref{sigma_11} then simplifies to
\begin{equation}\label{eqn: stress pz}
\sigma_{11} = -z\hat{C}_{11}^E(\theta)\partial^2_{x}{W} + \frac{\hat{e}_{31}(\theta)}{2d_p}V.
\end{equation}
Substituting Equation \eqref{eqn: stress elast} and \eqref{eqn: stress pz} into Equation \eqref{eqn:bending mom} and performing the integration gives the bending moment as 
\begin{equation}\label{eqn: bending mom result}
    M_{11}(x,t)=-B(\theta)\partial^2_{x}{W}+\eta(\theta) V,
\end{equation}
where $B(\theta)$ and $\eta(\theta)$ are given by
\begin{subequations}
\begin{align}
    &B(\theta) = \frac{d_0^3}{12}\hat{C}_{11}^0+2d_p\left(\frac{d_p^2}{3}+\frac{d_0d_p}{2}+\frac{d_0^2}{4}\right)\hat{C}_{11}^E(\theta),\label{eqn: B}\\
   &\eta(\theta) =\frac{1}{2}(d_0+d_p)\hat{e}_{31}(\theta).\label{eqn: eta} 
\end{align}
\end{subequations}
Substituting this expression for the bending moment (Eqn \eqref{eqn: bending mom result}) into Equation \eqref{eqn: motion} yields an equation relating the displacement $W$, voltage on the resistor $V$, and surface load $P$: 
\begin{equation}\label{eqn: bending equation}
    -B(\theta)\partial^4_{x}{W}+\eta(\theta)\partial^2_{x}{V}-I_b\partial^2_{t}{W}=-P.
\end{equation}

\subsubsection{Charge}
\noindent In order to derive the electrical equations for the system, we shall consider the electric displacement for the top plate. This is given by the constitutive relation (Eqn \eqref{eqn: topplate disp}) as
\begin{equation}
    D_3 = \hat{e}_{31}(\theta)\epsilon_{11} + \hat{\kappa}_{33}^{\varepsilon}(\theta)E_3,
\end{equation}
or
\begin{equation}\label{eqn: D3 inter}
    D_3 = -z\hat{e}_{31}(\theta)\partial^2_{x}{W} - \frac{\hat{\kappa}_{33}^{\varepsilon}(\theta)}{2d_p}V,
\end{equation}
using Equation \ref{eqn:strain relationship} and Equation \ref{eqn: voltage 1 bimorph} that relate the strain and electric field components to the out-of-plane displacement and voltage, respectively. In the literature \citep[e.g.,][]{renzi2016hydroelectromechanical,Erturk_Inman_2008,Erturk_Inman_2009}, it is common to take an expansion in $z$ about the midpoint $z=\frac{d_0+d_p}{2}$ of the top piezoelectric layer so that $D_3$ can be rewritten as
\begin{equation}
    D_3 = -\frac{1}{2}(d_0+d_p)\hat{e}_{31}(\theta)\partial^2_{x}{W} - \frac{\hat{\kappa}_{33}^{\varepsilon}(\theta)}{2d_p}V + O\left(z-\frac{d_0+d_p}{2}\right).
\end{equation}
This is equivalent to replacing the bending strain with the average bending strain in the piezoelectric layer. Because we assume that the plate is thin ($d_0+2d_p\ll L$, where $L$ is the half-length of the plate), the higher-order terms can be neglected. 

The average charge accumulated over each piezoelectric layer is given by \citep{renzi2016hydroelectromechanical}
\begin{equation}
    \hat{Q}=\int_{A}\boldsymbol{D}\cdot\boldsymbol{n}~\diff{A},
\end{equation}
where $A$ is the electrode area. Denoting the average electric charge per unit area as $Q(x,t)=\D{\hat{Q}}{A}=D_3$, we have
\begin{equation}\label{eqn: charge}
    Q(x,t)=-\eta(\theta)\partial^2_{x}{W} - C(\theta)V,
\end{equation}
where we have used Equation \eqref{eqn: eta} and denoted $C(\theta)=\frac{\hat{\kappa}_{33}^{\varepsilon}(\theta)}{2d_p}$ to be the electrical surface capacitance. Finally, Ohm's law gives
\begin{equation}
    \partial_t{Q}=GV,\label{eqn:ohms}
\end{equation}
where $G$ is the surface conductance, which is the reciprocal of the electrical resistance per unit area \cite{renzi2016hydroelectromechanical}.

Equations \eqref{eqn: bending equation}, \eqref{eqn: charge} and \eqref{eqn:ohms} fully describe the electrical and elastic dynamics of the bimorph in response to the pressure load $P$ of the fluid.

\subsubsection{Time harmonic formulation}
\noindent Suppose that the motion of the plate, the charge, the voltage, and the surface load are harmonic with frequency $\omega$. Specifically, we take
\begin{subequations}
\begin{align}
W(x,t)&=\Re\{\w(x)\e^{-\upi\omega t}\},\label{eqn: w harm}\\
Q(x,t)&=\Re\{\q(x)\e^{-\upi\omega t}\},\label{eqn: q harm}\\
V(x,t)&=\Re\{\vo(x)\e^{-\upi\omega t}\},\label{eqn: v harm}\\
P(x,t)&=\Re\{\p(x)\e^{-\upi\omega t}\}.\label{eqn: p harm}
\end{align}
\end{subequations}
where $\textrm{Re}$ denotes the real part, $t$ is time, and the scalar functions $\w$, $\q$, $\vo$, and $\p$ are complex-valued displacement, charge, voltage, and surface load. In the following, we omit $\Re$ for brevity. Substituting the above into Equation \eqref{eqn: bending equation}, \eqref{eqn: charge}, and \eqref{eqn:ohms} gives
\begin{subequations}
\begin{equation}\label{eqn: bending harmonic}
    -B(\theta)\partial^4_{x}{\w}+\eta(\theta)\partial^2_{x}{\vo}+I_b\omega^2\w=-\p,
\end{equation}
\begin{equation}\label{eqn: charge harm}
    \q(x)=-\eta(\theta)\partial^2_{x}{\w} - C(\theta)\vo,
\end{equation}
and
\begin{equation}\label{eqn: ohms harm}
    -\upi\omega q=G\vo,
\end{equation}
\end{subequations}
respectively. Substituting Equation \eqref{eqn: ohms harm} into Equation \eqref{eqn: charge harm} and rearranging for $\vo(x)$ gives
\begin{equation}\label{eqn:voltage}
    \vo(x)=\frac{\upi\omega\eta(\theta)}{G-\upi\omega C}\partial^2_{x}{\w}.
\end{equation}
Using this, Equation \eqref{eqn: bending harmonic} can then be written as
\begin{equation}
    \left[B(\theta)-\frac{\upi\omega\eta(\theta)^2}{G-\upi\omega C}\right]\partial^4_{x}{\w}-I_b\omega^2\w=\p,
\end{equation}
or equivalently
\begin{equation}\label{eom_plate_general}
    D\partial^4_{x}{\w}-I_b\omega^2\w=\p,
\end{equation}
where $D$ is given by
\begin{equation}\label{eqn: D param}
    D = B(\theta)+\frac{\omega\eta(\theta)^2}{\omega C+G\upi}.
\end{equation}
This result is similar to the one given by \citet{renzi2016hydroelectromechanical}, except Equation~\eqref{eqn: D param} is in dimensional form (see Equations \eqref{eqn: first mass moment}, \eqref{eqn: B} and \eqref{eqn: eta} for relevant expressions) and allows a rotation $\theta$ of the poling angle of the piezoelectric layers clockwise from the vertical. It should be noted that there is a typo in the last exponent of the expression for $B$ in \citet[][Eq.~(2.8)]{renzi2016hydroelectromechanical}.

\subsection{Equations of motion for the fluid}
\noindent For completeness, we present here the standard equations of linear water wave theory. The reader is referred to \citet{Linton2001} for further details. The problem is initially posed in a three-dimensional coordinate system with the $(x,y)$-plane coinciding with the undisturbed free surface of the fluid and the $z$-axis pointing vertically upwards. A flat seabed is situated at the plane $z=-H$, meaning the fluid occupies the region $\tilde{\Omega} =\mathbb{R}^2\times (-H,0)\setminus\tilde{\Gamma}$, where $\tilde{\Gamma}=(-L,L)\times\mathbb{R}\times\{-h\}$ is the region occupied by the infinitely thin plate, in which $2L$ is the extent of the plate in the $x$ direction. Note that the plate is parallel to the seabed, and can either be on the surface ($h=0$) or submerged ($-H<-h<0$).

Assuming an inviscid, incompressible fluid with irrotational motion implies the flow field $\mathbf{u}$ is of the form
\begin{equation}\label{potential1}
\mathbf{u}(x,y,z,t)=\nabla\Phi(x,y,z,t)\quad\text{for } (x,y,z)\in\tilde{\Omega},
\end{equation}
where $\Phi$, the velocity potential, satisfies
\begin{equation}\label{potential2}
\nabla^2\Phi=0\quad\text{for } (x,y,z)\in\tilde{\Omega},
\end{equation}
in which $\nabla=\begin{bmatrix}\partial_x &\partial_y&\partial_z\end{bmatrix}^\intercal$ and $\nabla^2=\nabla\cdot\nabla$. The linearised kinematic and dynamic free surface conditions are
\begin{subequations}\label{fs_conds_TD}
\begin{align}
\partial_t \zeta&=\partial_z\Phi&z=0,\\
\partial_t\Phi+g\zeta&=0&z=0, \label{eqn:fs_conds_pt2)}
\end{align}
\end{subequations}
respectively, where $\zeta(x,y,t)$ describes the free surface elevation. Equations \eqref{fs_conds_TD} hold for $x\in\mathbb{R}\setminus(-L,L)$ if the plate is on the surface or $x\in\mathbb{R}$ if the plate is submerged. The constant $g$ appearing in Equation \eqref{eqn:fs_conds_pt2)} is acceleration due to gravity. The boundary condition on the seabed is
\begin{equation}
\partial_z\Phi=0,
\end{equation}
for $z=-H$. We assume motion is time harmonic with angular frequency $\omega$ and invariant in the $y$ direction, implying
\begin{subequations}
\begin{align}
\Phi(x,y,z,t)&= \Re\left\{\phi(x,z)\mathrm{e}^{-\upi\omega t}\right\},\\
\zeta(x,y,t)&=\Re\left\{\hat\zeta(x)\e^{-\upi\omega t}\right\},
\end{align}
\end{subequations}
where $\phi$ is a complex-valued velocity potential satisfying
\begin{subequations}\label{fluid_no_plate}
\begin{align}
\Delta\phi&=0&\text{for}\quad (x,z)\in\Omega,\label{Laplace_equation}\\
\partial_z\phi &= \alpha \phi& \text{for}\quad z=0,\label{fs_cond}\\
\partial_z\phi&=0&\text{for}\quad z=-H,
\end{align}
where $\Delta\equiv\partial_x^2+\partial_z^2$ is the two-dimensional Laplacian operator and $\alpha=\omega^2/g$. The two-dimensional fluid domain is given by $\Omega=(\mathbb{R}\times(-H,0))\setminus\Gamma$, where the region occupied by the plate's cross section is $\Gamma = (-L,L)\times\{-h\}$. Note that following from Equations \eqref{fs_conds_TD}, Equation \eqref{fs_cond} holds for $x\in\mathbb{R}\setminus(-L,L)$ if the plate is on the surface or $x\in\mathbb{R}$ if the plate is submerged. In the far-field the solution is subject to the Sommerfeld radiation condition. This is given by
\begin{equation}\label{eqn:Sommerfeld}
\left\{\partial_{x}\phi^{\rm sc}\mp\upi k\phi^{\rm sc}\right\}\to 0\quad\text{as}\quad x\to\pm\infty,
\end{equation}
\end{subequations}
where $k$ is the positive real solution of the dispersion equation
\begin{equation}
    k\tanh(k H)=\frac{\omega^2}{g},
\end{equation}
$\phi^{\rm sc}=\phi - \phi^{\rm in}$ is the scattered wave potential, and 
\begin{equation}\label{eqn:plane_inc}
\phi^{\rm in}(x,z)=\frac{-\upi g A}{\omega\cosh kH} \mathrm{e}^{\upi k x}\cosh k(z+H),
\end{equation}
is the velocity potential of a plane incident wave with amplitude $A$. We take $A=1$\,m throughout this paper. To complete the system of equations \eqref{fluid_no_plate}, it remains to specify the boundary conditions on the plate.

\subsection{The coupled system}\label{sec: coupled eqns of motion}
\noindent The complex velocity potential, $\phi$, is coupled to the complex plate displacement function, $\w$, through linearised kinematic and dynamic conditions, respectively. The kinematic condition is given by
\begin{equation}
    \partial_z \phi=-\upi \omega \w\quad\text{for}\quad (x,z)\in\Gamma,
\end{equation}
which enforces that the fluid and plate velocities coincide at their interface. The dynamic condition (restated here from Equation \eqref{eom_plate_general}), is
\begin{equation*}
    D\partial_x^4\w - I_b \omega^2\w = \p \quad\text{for}\quad (x,z)\in\Gamma,
\end{equation*}
where the loading $\p$ is now determined from the fluid pressure as 
\begin{equation}\label{hydrodyn_load}
    \p=\begin{cases}
        \upi\omega\rho_w\phi- \rho_w g\w,&h=0,\\
        -\upi\omega\rho_w\llbracket\phi\rrbracket,&0<h<H,
    \end{cases}
\end{equation}
where $\rho_w$ is the density of the fluid and $\llbracket\phi\rrbracket=\phi_+-\phi_-$ denotes the jump of $\phi$ across the plate. The notation $\phi_+$ and $\phi_-$ denotes the limit of $\phi$ as $z\rightarrow-h$ from above and below, respectively. Note that for the plate on the surface ($h=0$), the first and second summands contributing to $\p$ are the hydrodynamic and hydrostatic pressure, respectively. Finally, we have that the jump in the normal derivative of the potential across the submerged plate is zero, namely,
\begin{equation}
\llbracket\partial_z\phi\rrbracket=0\quad\text{for}\quad (x,z)\in\Gamma.
\end{equation}
For the boundary conditions on the plate, we assume that the edges are either clamped:
\begin{equation}\label{eqn:clamped}
 \w = \partial_x \w = 0\quad\text{for }x=\pm L,
\end{equation}
or simply supported:
\begin{equation}\label{eqn:simply-supported}
 \w = \partial_x^2 \w = 0\quad\text{for }x=\pm L.
\end{equation}
In summary, we have the following coupled system:
\begin{equation}\label{eqn: coupled system}
\begin{aligned}
\Delta\phi&=0 & (x,z)\in\Omega,\\
\partial_z\phi&=0 & (x,z)\in\mathbb{R}\times\{H\},\\
\partial_z\phi &= \frac{\omega^2}{g}\phi & (x,z)\in(\mathbb{R}\times\{0\})\setminus\Gamma,\\
\partial_z \phi&=-\upi \omega \w & (x,z)\in\Gamma,\\
D\partial_x^4\w - I_b \omega^2\w &= \p & (x,z)\in\Gamma,\\
\llbracket\partial_z\phi\rrbracket&=0 & (x,z)\in\Gamma,\\
\left\{\partial_{x}\phi^{\rm sc}\mp\upi k\phi^{\rm sc}\right\}&\to 0 &  x\to\pm\infty,
\end{aligned}
\end{equation}
with either clamped (Eqn \eqref{eqn:clamped}) or simply supported (Eqn \eqref{eqn:simply-supported}) boundary conditions.
\begin{figure}
\begin{center}
\begin{tikzpicture}[scale=2]
\draw[dashed] (1.5,1) -- (1.5,3);
\draw[dashed] (4.5,1) -- (4.5,3);
\draw[dashed] (0,2) -- (6,2);
\draw (0,1) -- (6,1) node[above,xshift = -165] {$\partial_z\phi = 0$};
\draw[line width=1mm, black] (1.5,2) -- (4.5,2) node[above,xshift = -85] {$D \partial_x^4  \w   - I_b \omega^2 \: \w  = \p $};
\node[text width=3cm] at (6.8,3.02) {$z = 0$};
\node[text width=3cm] at (6.8,2.02) {$z = -h$};
\node[text width=3cm] at (6.8,1.02) {$z = -H$};
\draw[solid, blue] (0,3) -- (6,3) node[above,xshift = -125] {};
\node[text width=3cm] at (1.75,3.30) {$\partial_{z}\phi=\frac{\omega^2}{g} \phi$};
\draw[line width=0.5mm, black,->] (3,3) -- (3,3.5) node[right] {$z$};
\draw[line width=0.5mm, black,->] (3,3) -- (3.5,3) node[above] {$x$};
\node[text width=3cm] at (1,1.5){$\displaystyle{\Delta\phi = 0}$};
\node[text width=6cm] at (4,1.75){-$\upi\omega \w  = \partial_z\phi $};
\node[text width=2cm] at (1.65,0.8){$x=-L$};
\node[text width=2cm] at (4.75,0.8){$x=L$};
\end{tikzpicture}
\end{center}
\caption{Visualisation of the problem setup. For the plate on the surface ($h=0$), the complex-valued surface load is given by $\p = \upi\omega\rho_w\phi-\w \rho_w g$. When the plate is submerged ($-H<h<0$), the surface load is instead $\p=-\upi\omega\rho_w\llbracket\phi\rrbracket$.}
\label{fig:schematic}
\end{figure}
Figure \ref{fig:schematic} shows a schematic of the above equations.

\section{Solution methods}\label{solution_sec}

\subsection{Surface problem}
\noindent For the plate floating on the ocean's surface we use the approach described by \citet{Meylan_Challis_Thamwattana_Wegert_Wilks_2025}, in which an expansion in modes and a Green’s function are used to analyse the behaviour. We refer the reader to the referenced work for further details.

\subsection{Submerged problem}
\noindent In the following, we describe our new approach for solving the problem of a submerged plate by decomposing into a diffraction problem (in which the incident wave scatters off a fixed plate) and radiation problems (in which the fluid motion is forced by plate vibration in a given mode). We then obtain the solution of the coupled problem (Eqns \eqref{eqn: coupled system}) using the modal expansion method of \citet{newman1994wave}. The component diffraction and radiation problems are solved by formulating them as hypersingular boundary integral equations and solving numerically using a constant panel method. Our method extends earlier work of \citet{Meylan_Challis_Thamwattana_Wegert_Wilks_2025} for submerged plates. Our method is advantageous compared to the eigenfunction expansion method used by \citet{renzi2016hydroelectromechanical} because it does not require the solution of a complicated dispersion relation. Furthermore, the method extends to non-horizontal plates and to two-dimensional plates of arbitrary shape, however it becomes inefficient as the plate becomes larger.

\subsubsection{Diffraction problem}\label{diffraction_sec}
\noindent The velocity potential associated with diffraction is of the form $\phi=\phi^{\mathrm{in}}+\phi^{\mathrm{di}}$, where $\phi^{\mathrm{di}}$ is the as yet unknown diffraction potential. Assuming that the plate is held fixed, we have $\partial_z(\phi^{\mathrm{in}}+\phi^{\mathrm{di}})=0$ for $(x,z)\in\Gamma$. Since $\phi$ also satisfies Equations \eqref{fluid_no_plate} and $\phi^{\rm di}$ satisfies the Sommerfeld radiation condition (Eqn \eqref{eqn:Sommerfeld}), we obtain the following system of equations for $\phi^{\rm di}$
\begin{subequations}
\begin{align}
\Delta \phi^{\rm di}&=0&(x,z)\in\Omega,\\
\partial_z\phi^{\rm di}&=\alpha\phi^{\rm di}&z=0,\\
\partial_z\phi^{\rm di}&=-\partial_z\phi^{\rm in}&(x,z)\in\Gamma\label{diffraction_forcing},\\
\partial_z\phi^{\rm di}&=0&z=-H,\\
(\partial_{|x|}-\upi k)\phi^{\mathrm{di}}&\to 0&\text{as } |x|\to\infty.
\end{align}
\end{subequations}

The Green's function for a two dimensional fluid of finite depth is \citep{Linton2001}
\begin{align}\label{eqn:Green's function}
2\pi G(x,z;x^\prime,z^\prime)&=\ln r-\ln \tilde{r}\nonumber\\ &-2\Uint_0^\infty
\frac{\cos \mu X}{\cosh \mu h}
\left[\frac{\cosh\mu(z+H)\cosh\mu(z^\prime+H)}{\mu\sinh\mu H-\alpha\cosh \mu H}+\frac{\e^{-\mu H}}{\mu}\sinh\mu z\sinh \mu z^\prime\right]\upd\mu,
\end{align}
in which $X=x-x^\prime$, $r=\sqrt{(x-x^\prime)^2+(z-z^\prime)^2}$ and $\tilde{r}=\sqrt{(x-x^\prime)^2+(z+z^\prime)^2}$, which satisfies
\begin{equation}
\begin{aligned}
\Delta G&=\delta(x-x^\prime)\delta(z-z^\prime)&(x,z)\in\Omega\cup\Gamma,\\
\partial_z G&=\alpha G&z=0,\\
\partial_z G&=0&z=-H,\\
(\partial_{|x|}-\upi k)G&\to 0&\text{as } |x|\to\infty.
\end{aligned}
\end{equation}
The method to numerically calculate the integral in the Green's function expression is discussed in \ref{greens_fn_appendix}. 
Note that $G$ can be represented as the sum of a singular part and a regular part as
\begin{equation}
2\pi G(x,y;x^\prime,y^\prime)=\ln r+M(x,y;x^\prime,y^\prime).
\end{equation}
Applying Green's third identity to $\phi^{\rm di}$ gives
\begin{equation}
\phi^{\mathrm{di}}(x,z)=-\int_{-L}^L \partial_{z^\prime}G(x,z;x^\prime,-h)\llbracket\phi^{\mathrm{di}}(x^\prime)\rrbracket\upd x^\prime.
\end{equation}
The incident forcing condition (Eqn \eqref{diffraction_forcing}) yields
\begin{equation}
-\partial_z\phi^{\mathrm{in}}(x,-h)=-\partial_z\int_{-L}^L \partial_{z^\prime}G(x,-h;x^\prime,-h)\llbracket\phi^{\mathrm{di}}(x^\prime)\rrbracket\upd x^\prime.\label{int_eq_1}
\end{equation}
Interchanging the derivative and the integral in the above expression is permitted provided the integral is interpreted as a Hadamard finite part integral \citep[here denoted $\Xint$, cf.][]{PARSONS1992313}. The Hadamard finite part integral is defined for a continuous function $f:[a,b]\to\mathbb{C}$ as
\begin{equation}
\Xint_a^b \frac{f(t)}{(t-s)^2}\upd t=\lim_{\varepsilon\to 0^+}\left\{\int_a^{s-\varepsilon}\frac{f(t)}{(t-s)^2}\upd t+\int_{s+\varepsilon}^b\frac{f(t)}{(t-s)^2}\upd t-\frac{2f(s)}{\varepsilon}\right\}.
\end{equation}
Equation \eqref{int_eq_1} becomes
\begin{align}\label{int_eq_2}
-\partial_z\phi^{\mathrm{in}}(x,-h)=-\Xint_{-L}^L \partial_z\partial_{z^\prime}G(x,-h;x^\prime,-h)\llbracket\phi^{\mathrm{di}}(x^\prime)\rrbracket\upd x^\prime.
\end{align}
Our goal is to discretise the integral equation above using a constant panel method. We divide the interval $[-L,L]$ into $N$ subintervals of equal length with midpoints $x_1,\dots,x_N$ and spacing $x_{j+1}-x_j=\Delta x$. Equation \eqref{int_eq_2} becomes
\begin{subequations}
\begin{align}
-\partial_z\phi^{\mathrm{in}}(x,-h)&=-\sum_{j=1}^N\Xint_{x_j-\Delta x/2}^{x_j+\Delta x/2} \partial_z\partial_{z^\prime}G(x,-h;x^\prime,-h)\llbracket\phi^{\mathrm{di}}(x^\prime)\rrbracket\upd x^\prime\\
&\approx-\sum_{j=1}^N\llbracket\phi^{\mathrm{di}}(x_j)\rrbracket\Xint_{x_j-\Delta x/2}^{x_j+\Delta x/2} \partial_z\partial_{z^\prime}G(x,-h;x^\prime,-h)\upd x^\prime,\label{constant_panel}
\end{align}
\end{subequations}
where we have assumed $\llbracket\phi^{\mathrm{di}}(x)\rrbracket$ is slowly varying with respect to $\Delta x$ in order to apply the constant panel approximation. Evaluating both sides of Equation \eqref{constant_panel} at the collocation points $x_i$ yields the matrix equation
\begin{equation}
-\sum_{j=1}^N K_{ij}\llbracket\phi^{\mathrm{di}}(x_j)\rrbracket\approx-\partial_z\phi^{\mathrm{in}}(x_i,-h),
\end{equation}
where the terms $K_{ij}$ are introduced as approximations of the integrals in Equation \eqref{constant_panel}, i.e.
\begin{equation}
K_{ij}\approx\Xint_{x_j-\Delta x/2}^{x_j+\Delta x/2} \partial_z\partial_{z^\prime}G(x,-h;x^\prime,-h)\upd x^\prime.
\end{equation}
Our strategy to approximate these terms is to integrate the singular part exactly (in the Hadamard finite part sense) and approximate the integral of the regular part using the midpoint rule. We first compute
\begin{equation}
2\pi\partial_z\partial_{z^\prime}G(x,y;x^\prime,y^\prime)\biggr\rvert_{z=z^\prime} = \frac{-1}{(x-x^\prime)^2}+\partial_z\partial_z^\prime M(x,y;x^\prime,y^\prime).
\end{equation}
Then the terms $K_{ij}$ are computed as
\begin{equation}
2\pi K_{ij} = \begin{cases}
  \displaystyle  \frac{4}{\Delta x}+\partial_z\partial_{z^\prime} M(x_i,-h;x_i,-h)\Delta x&i=j\\
  \displaystyle  \frac{1}{x_i-x_j+\Delta x/2}-\frac{1}{x_i-x_j-\Delta x/2}+\partial_z\partial_{z^\prime} M(x_i,-h;x_j,-h)\Delta x&i\neq j.
\end{cases} 
\end{equation}

\subsubsection{Modes of vibration of an elastic beam}
\noindent The equations of motion of a uniform elastic beam satisfy the eigenvalue problem
\begin{equation}\label{eigenvalue_beam}
D\partial_x^4 \w_n=I_b\omega_n^2 \w_n,
\end{equation}
for $x\in(-L,L)$, where $\omega_n$ and $\w_n$ are the angular frequencies and modes of vibration, respectively. We further introduce the quantities $\kappa_n=\sqrt[^4]{I_b \omega_n^2/D}$. The eigenvalue problem Equation \eqref{eigenvalue_beam} is solved in tandem with boundary conditions at $x=\pm L$. The modes of vibration satisfy the prescribed boundary conditions at $x=\pm L$, namely, those for either clamped (Eqn \eqref{eqn:clamped}) or simply supported (Eqn \eqref{eqn:simply-supported}) plates.
Note that the eigenmodes are scaled to be orthonormal, namely
\begin{equation}\label{orthonormality}
    \int_{-L}^L \w_n(x){\w_m(x)}\upd x = \delta_{mn},
\end{equation}
where $\delta_{mn}$ is the Kronecker delta. The expressions for the eigenmodes and eigenvalues can be found in \citet[Chapt. 7.3.,][]{Reddy_2006}.

\subsubsection{Radiation problem}
\noindent We consider the radiation problems, whose solutions are the complex velocity potentials $\phi_n$ resulting from excitation by forced motion of the plate in a single mode of vibration $\w_n$, in the absence of an incident wave. The boundary value problem to be solved is
\begin{equation}
\begin{aligned}
\Delta \phi_n&=0&(x,z)\in\Omega,\\
\partial_z\phi_n&=\alpha\phi_n&z=0,\\
\partial_z\phi_n&=-\upi\omega \w_n&(x,z)\in\Gamma,\\
\partial_z\phi_n&=0&z=-H,\\
(\partial_{|x|}-\upi k)\phi_n&\to 0&\text{as } |x|\to\infty.
\end{aligned} 
\end{equation}
Following the steps outlined in \textsection\ref{diffraction_sec}, we find
\begin{equation}
\phi_n(x,z)=-\int_{-L}^L \partial_{z^\prime}G(x,z;x^\prime,-h)\llbracket\phi_n(x^\prime)\rrbracket\upd x^\prime.
\end{equation}
Accordingly, the pressure difference $\llbracket\phi_n\rrbracket$ can be computed numerically at the collocation points $x_j$ by solving the linear system of equations
\begin{equation}
-\sum_{j=1}^N K_{ij}\llbracket\phi_n(x_j)\rrbracket\approx-\upi\omega \w_n(x_i).
\end{equation}

\subsubsection{Scattering by a piezoelectric plate}
\noindent The coupled problem Equations \eqref{eqn: coupled system} is solved using a dry modes expansion \cite{newman1994wave} of the form
\begin{subequations}
\begin{align}
    \phi&=\phi^{\mathrm{in}}+\phi^{\mathrm{di}}+\sum_{n}c_n\phi_n,\\
    \w&=\sum_n c_n\w_n,
\end{align}
\end{subequations}
where the coefficients $c_n$ must be determined. Substituting these into the dynamic condition of the plate (Eqn \eqref{eom_plate_general}) with hydrodynamic loading given by Equation \eqref{hydrodyn_load} gives
\begin{equation}\label{dry_modes_eq_1}
D\sum_{n=1}^\infty c_n \kappa_n^4 \w_n -\omega^2I_b\sum_{n=1}^\infty c_n\w_n+\upi\omega\rho\sum_{n=1}^\infty c_n\llbracket\phi_n\rrbracket=-\upi\omega\rho\llbracket\phi^{\mathrm{di}}\rrbracket,
\end{equation}
where we have used the fact that $\llbracket\phi^{\mathrm{in}}\rrbracket=0$. A linear system of equations for the unknown coefficients $c_n$ is obtained by truncating the sum in Equation \eqref{dry_modes_eq_1} after $N_m$ terms, then taking the $\mathrm{L}^2(-L,L)$ inner product of both sides of the resulting equation with $\w_m$ for $m=1,\dots,N_m$. We obtain
\begin{equation}
D\kappa_m^4 c_m-\omega^2I_b c_m + \upi\omega\rho_w\sum_{n=1}^{N_m} \left(\int_{-L}^L\llbracket\phi_n(x)\rrbracket\overline{\w_m(x)}\upd x\right) c_n=-\upi\omega\rho_w\int_{-L}^L\llbracket\phi^{\mathrm{di}}(x)\rrbracket\overline{\w_m(x)}\upd x,
\end{equation}
where we have used Equation \eqref{orthonormality}. In matrix form, we have
\begin{equation}
(\boldsymbol{K}_{\rm stiffness}-\omega^2\boldsymbol{M}_{\rm mass}-\omega^2\boldsymbol{A}_{\rm mass}-\upi\omega \boldsymbol{B}_{\rm damping})\mathbf{c}=\mathbf{f},
\end{equation}
where $\boldsymbol{K}_{\rm stiffness}$, $\boldsymbol{M}_{\rm mass}$, $\boldsymbol{A}_{\rm mass}$ and $\boldsymbol{B}_{\rm damping}$ are the $N_m\times N_m$ stiffness, mass, added mass and damping matrices given by
\begin{subequations}
\begin{align}
(\boldsymbol{K}_{\rm stiffness})_{mn} &= D\kappa_m^4\delta_{mn},\\
(\boldsymbol{M}_{\rm mass})_{mn} &=I_b\delta_{mn},\\
(\boldsymbol{A}_{\rm mass})_{mn}&=-\frac{1}{\omega^2}\Re\left\{\upi\omega\rho\int_{-L}^L\llbracket\phi_n(x)\rrbracket\overline{\w_m(x)}\upd x\right\},\\
(\boldsymbol{B}_{\rm damping})_{mn} &=-\frac{1}{\omega}\Im\left\{\upi\omega\rho\int_{-L}^L\llbracket\phi_n(x)\rrbracket\overline{\w_m(x)}\upd x\right\},
\end{align}
the vector of unknown coefficients is given by $\mathbf{c}=[c_1,\dots,c_{N_m}]^\intercal$ and the forcing vector is given by
\begin{equation}
(\mathbf{f})_m=-\upi\omega\rho_w\int_{-L}^L\llbracket\phi^{\mathrm{di}}(x)\rrbracket\overline{\w_m(x)}\upd x.
\end{equation}
\end{subequations}

\subsubsection{Reflection and transmission}
\noindent The velocity potential of the hydrodynamic coupled problem is
\begin{subequations}
\begin{align}
\phi&=\phi^{\rm in}+\phi^{\rm di} + \sum_nc_n\phi_n^{\rm ra},\\
&=\phi^{\rm in}+\phi^{\rm sc},
\end{align}
\end{subequations}
where
\begin{equation}\label{Phi_sc_def}
\phi^{\rm sc}(x,z)=-\int_{-L}^L \partial_{z^\prime}G(x,z;x^\prime,-h)\llbracket\phi(x^\prime)\rrbracket\upd x^\prime.
\end{equation}
Asymptotically, we have
\begin{equation}\label{refl_trans}
\phi(x,z)\sim\begin{cases}-\frac{\upi g A}{\omega\cosh kH}\cosh(k(z+H)) (\e^{\upi kx}+R\e^{-\upi k x})&\mbox{as }x\to-\infty,\\
-\frac{\upi g A}{\omega\cosh kH}\cosh(k(z+H)) T\e^{\upi kx}&\mbox{as }x\to\infty,
\end{cases}
\end{equation}
where $R$ and $T$ are the reflection and transmission coefficients, respectively. The asymptotic behavior of the Green's function is given by \cite{Linton2001}
\begin{equation}
\partial_zG(x,z;x^\prime,-h)\sim \frac{-\upi}{H(1+\sinh(2kH)/2kH)}\cosh(k(z+H))\sinh(k(H-h))\e^{\upi k|x-x'|},
\end{equation}
as $|x-x'|\to\infty$. Equating Equation \eqref{Phi_sc_def} with Equation \eqref{refl_trans} using the above leads to
\begin{subequations}
\begin{align}
R &= -\frac{\omega\cosh kH}{AgH(1+\sinh(2kH)/2kH)}\sinh(k(H-h))\int_{-L}^L\e^{\upi k x^\prime}\llbracket\phi(x^\prime)\rrbracket\upd x^\prime,\\
T&=1-\frac{\omega\cosh kH}{AgH(1+\sinh(2kH)/2kH)}\sinh(k(H-h))\int_{-L}^L\e^{-\upi k x^\prime}\llbracket\phi(x^\prime)\rrbracket\upd x^\prime.
\end{align}
\end{subequations}
The proportion of energy absorbed can be calculated using the reflection and transmission coefficients as
\begin{equation}\label{eqn: energy absorb}
    1-\lvert R\rvert^2 -\lvert T\rvert^2.
\end{equation}

\change{\subsubsection{Calculation of energy absorption efficiency}}

\noindent\change{We need to consider a realistic wave spectrum to talk about efficiency for realistic conditions. We assume here the Pierson–Moskowitz spectrum although other spectra could be considered. The energy absorption efficiency is}
\change{\begin{equation}\label{eqn: eff}
    E_{\mathrm{eff}}=\frac{E_p}{E}=\frac{\int_{0}^{\infty}S(\omega)\left(1-\lvert R\rvert^2 -\lvert T\rvert^2\right) ~\mathrm{d}\omega}{\int_{0}^{\infty} S(\omega)~\mathrm{d}\omega},
\end{equation}}\change{where $S(\omega)$ is given by the Pierson–Moskowitz spectrum \cite{pierson1964proposed,dean1991water}}
\change{\begin{equation}
    S(\omega)
=
a g^{2}\,\omega^{-5}
\exp\!\left(
-\frac{5}{4}\left(\frac{2\pi}{T_p\omega}\right)^4
\right)
\end{equation}}\change{with $a = 8.1\times 10^{-3}$. This spectrum describes the distribution of energy as a function of frequency for a wave with peak period $T_p$, and $E_p$ is the energy absorbed by the PWEC for the given peak period. To approximate Equation~\eqref{eqn: eff} we truncate the integrals so that they are between the minimum and maximum frequencies considered in the results.} \changerevtwo{Note that Equation~\eqref{eqn: eff} can be calculated using either the transmission and reflection coefficients (Eqn.~\ref{eqn: energy absorb}) or the nearfield power (see \ref{sec:val coe}).}

\subsubsection{Validation using conservation of energy}\label{sec:coe}
\noindent We check that the far-field power takeoff given by hydrodynamics for both the submerged plate and floating plate matches the near-field power takeoff that can be calculated using the equations for piezoelectricity \citep{renzi2016hydroelectromechanical}. The far-field power takeoff is given by
\begin{equation}
    P_{\mathrm{farfield}}=\frac{1}{2}\rho_wgA^2C_g\left(1-\lvert R\rvert^2 -\lvert T\rvert^2\right),
\end{equation}
where the group velocity $C_g$ is given by
\begin{equation}
C_g = \frac{\omega}{2k}\left(1+\frac{2kH}{\sinh(2kH)}\right). 
\end{equation}
The near-field power takeoff is given by
\begin{equation}\label{eqn: nearfield power}
    P_{\mathrm{nearfield}} = \frac{G}{2}\int_{-L}^L\lvert\vo\rvert^2~\mathrm{d}x =\frac{\omega^2\eta^2 G}{2(G^2+\omega^2C^2)}\int_{-L}^L\left\lvert\partial_x^2\w\right\rvert^2~\mathrm{d}x.
\end{equation}
In our numerical experiments, we found that these two expressions for the power takeoff generally matched to a relative precision of $10^{-13}$. \changerevtwo{\ref{sec:val coe} shows the validation for a simply supported bimorph made of PZT-5H.}

\subsubsection{Validation using other numerical methods}
Our method has been validated using both boundary element code \citep{Vipin_Koley_2022} and eigenfunction matching code \citep{meylan2009water}. A comparison with the solution by eigenfunction matching is given in \ref{comparison} for a purely elastic plate. This also serves as a benchmark calculation.

\section{Numerical studies}\label{numerics_sec}

\subsection{Parameter values and single-frequency results} \label{sec: initial results}
\noindent In our numerical experiments, we take the depth of the fluid to be $H=10$\,m, the total length of the plate to be $2L=20$\,m, and the depth of submergence to be $h=2$\,m for the submerged plate. For the elastic substrate, we use the material values for silicone rubber with a Young's modulus of $E_0=3.2\times10^6\,\textrm{Pa}$, a Poisson's ratio of $\nu_0=0.49$, and a density of $\rho_0=1250\,\textrm{kg\,m}^{-3}$ \citep{tanaka_experimental_2015}. Under plane stress, the resulting stiffness of the elastic substrate is then $\hat{C}_{11}^0=\frac{E_0}{1-\nu_0^2}\approx4.21\times10^6\,\textrm{Pa}$. We take the thickness of the substrate to be $d_0=0.01$\,m.
\begin{table}[t]
    \centering
    \caption{Piezoelectric material values for vertically poled PVDF and PZT-5H. For PVDF, we compute the above quantities using $\hat{C}_{11}=E_p/(1-\nu_p^2)$ and $\hat{e}_{31}=d_{31}E_p/(1-\nu_p)$ where $E_p=3\times10^9\,\mathrm{Pa}$ and $d_{31}=2.3\times10^{-12}\,\mathrm{mV}^{-1}$ \citep{MeasurementSpecialities}, and $\nu_p=0.392$ \citep{Nalwa_1995}. For PZT-5H, we use the expressions in \ref{appendix1} for $\theta=0$.}
    \label{tab:material values}
    \begin{tabular}{c|ccccc}
        Material & $\rho_p$ ($\textrm{kg\,m}^{-3}$) & $\hat{C}_{11}$ (Pa) & $\hat{e}_{31}$ ($\textrm{C\,m}^{-2}$) & $\hat{\kappa}^\varepsilon_{33}$ ($\textrm{F\,m}^{-1}$) & Source \\\hline
        PVDF & 1780 & $3.54\times10^9$ & 0.113 & $1.13\times10^{-10}$ & \citep{Nalwa_1995,MeasurementSpecialities} \\
        PZT-5H & 7500 & $6.55\times10^{10}$ & -23.2 & $1.77\times10^{-8}$ & \citep{YangIntroPZ}
    \end{tabular}
\end{table}
For the top and bottom piezoelectric plates, we use the material properties given in Table~\ref{tab:material values}\change{ which are based on known engineering values for PVDF and PZT-5H}. We initially fix the ratio of $G/C=1$\,s$^{-1}$, where we recall $G$ is the surface conductance and $C$ is the electrical surface capacitance. It is important to note that the surface conductance is a tunable parameter in the model related to the resistivity, and we will vary this ratio of $G/C$ in later numerical experiments. We take the thickness of the piezoelectric plates to be $d_p=1.1\times10^{-4}\,\mathrm{m}$. Finally, unless otherwise stated, we take $\Delta x=0.05~\mathrm{m}$. \change{We note that the geometric quantities chosen in this work are the same as those used by \citet{renzi2016hydroelectromechanical} and are similar to those that appear in \citet{Zheng_etal2021}. We believe that these are representative of realistic engineering scales; however, further experimental work is needed to understand the engineering practicality.}

\begin{figure}[p]
    \centering
    \begin{subfigure}{\textwidth}
        \centering
        \includegraphics[width=0.98\linewidth]{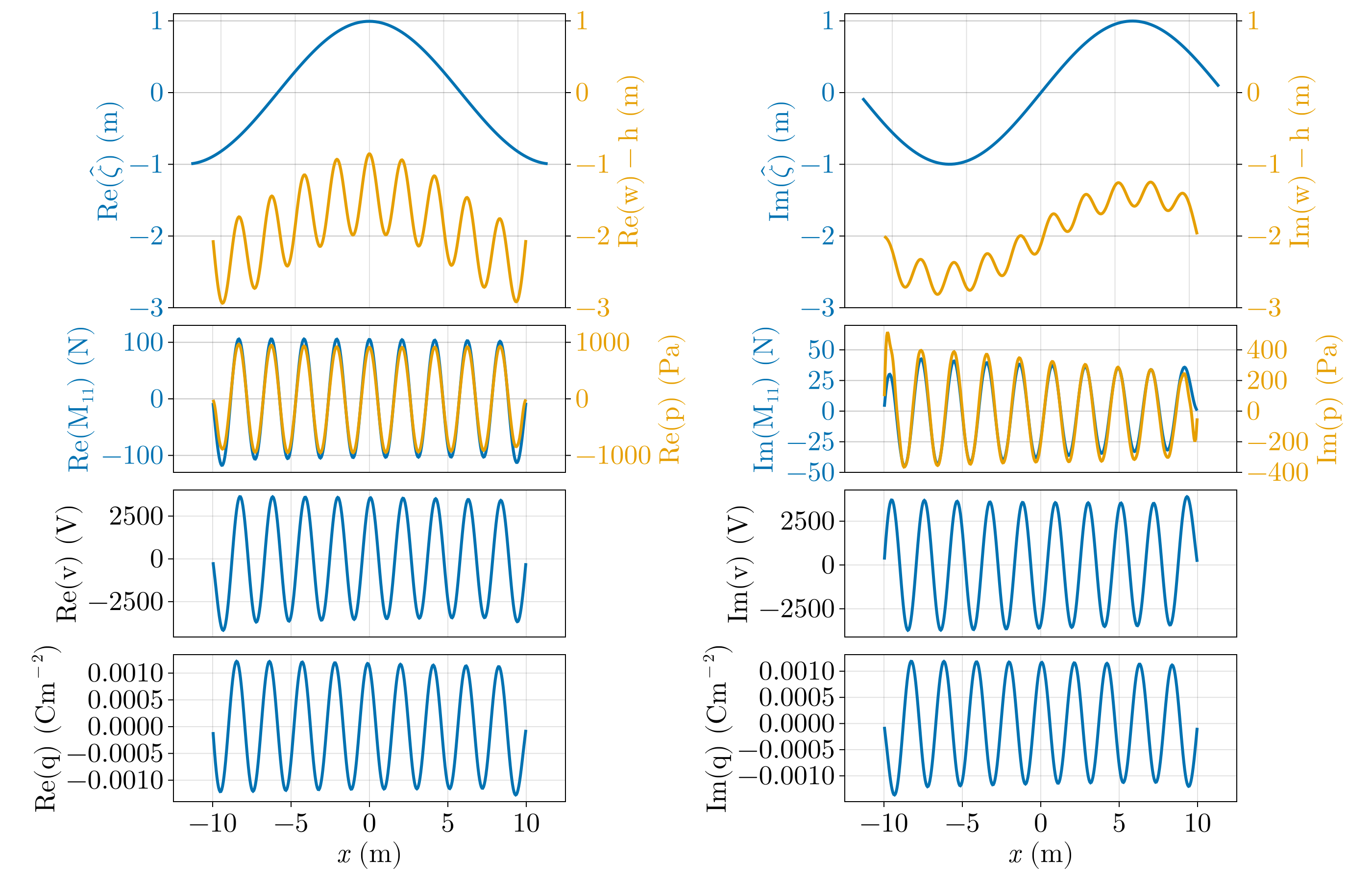}
        \caption{PVDF}
        \label{fig:single-freq-PVDF}
    \end{subfigure}
    \begin{subfigure}{\textwidth}
        \centering
        \includegraphics[width=0.98\linewidth]{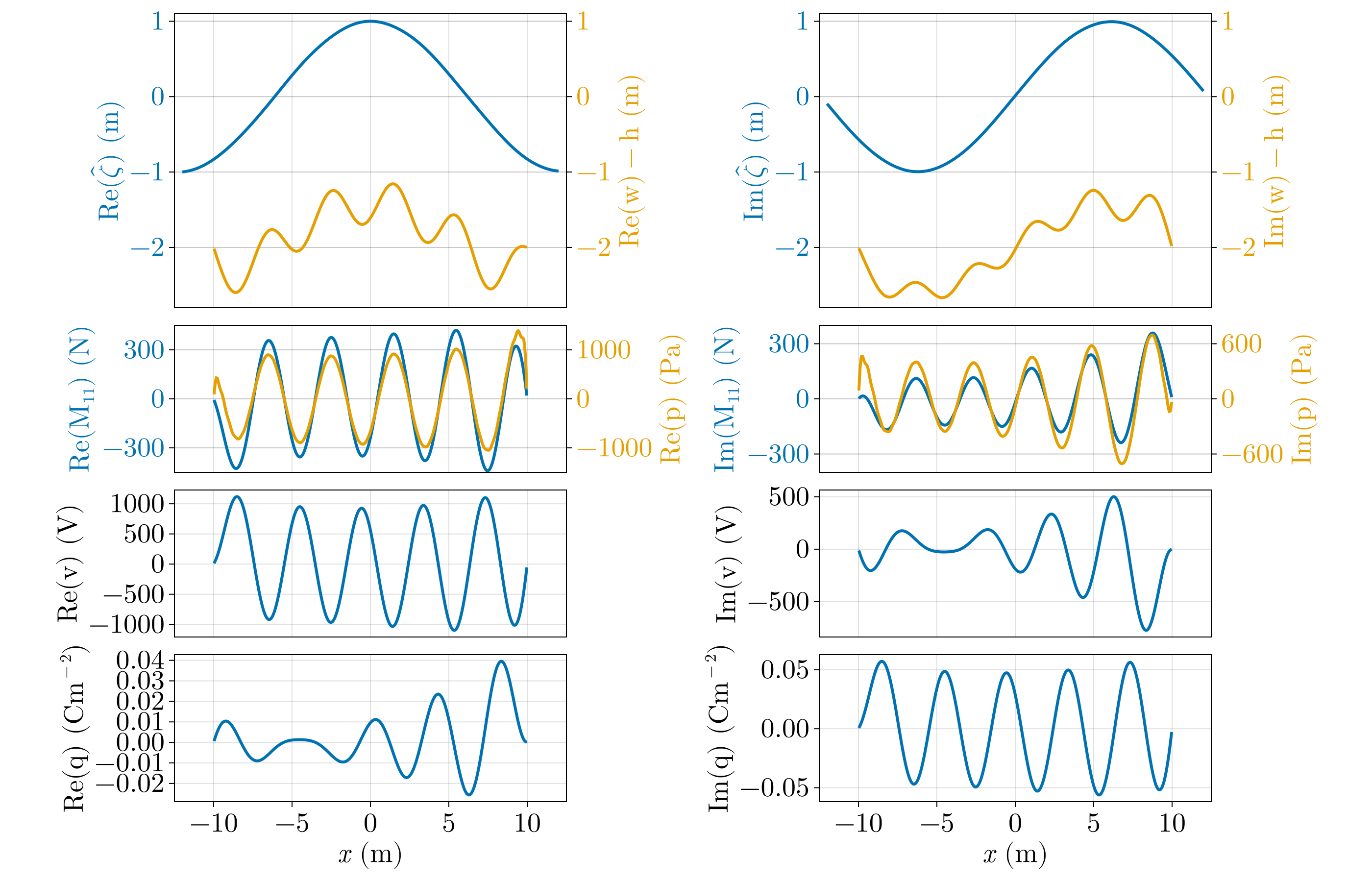}
        \caption{PZT-5H}
        \label{fig:single-freq-PZT5H}
    \end{subfigure}
    \caption{Single-frequency real and imaginary results for the free-surface displacement, plate displacement, moment, pressure, voltage, and charge for a submerged simply-supported bimorph made of \change{(a) PVDF and (b) PZT-5H} for a wave period of 4 seconds. \changerevtwo{The blue and yellow lines correspond to the coloured y-axis labels.}}
    \label{fig:single-freq}
\end{figure}

Before considering the power conversion across a range of frequencies, we briefly consider single-frequency results for the real and imaginary parts of the free-surface displacement, plate displacement, moment, pressure, voltage, and charge for the submerged simply-supported bimorph. Figures~\ref{fig:single-freq-PZT5H} and \ref{fig:single-freq-PVDF} show these quantities for a wave period of 4 seconds for submerged bimorphs made of PZT-5H and PVDF, respectively. The results for the voltage and charge demonstrate Ohm's law in the frequency domain \eqref{eqn: ohms harm}, in particular that the charge is proportional to a 90$^\circ$ phase shift of the voltage (cf. $\Re(\vo)$ to $\Im(\q)$). In addition, the increase in the number of oscillations in the plate displacement for PVDF compared to PZT-5H demonstrates the increased stiffness of the latter material. Finally, it should be noted that the plate displacement, moment, and pressure are in-phase, while the displacement is out of phase with the voltage and charge. The latter is because these quantities are related to the displacement through higher order derivatives, while the former is because of the relationship between displacement, moment, and pressure given by Equation~\eqref{eqn: motion} and Equation~\eqref{eqn: bending mom result}.

\subsection{Surface plate versus submerged plate}\label{sec: surf vs sub}

\begin{figure}[!t]
    \centering
    \includegraphics[width=\linewidth]{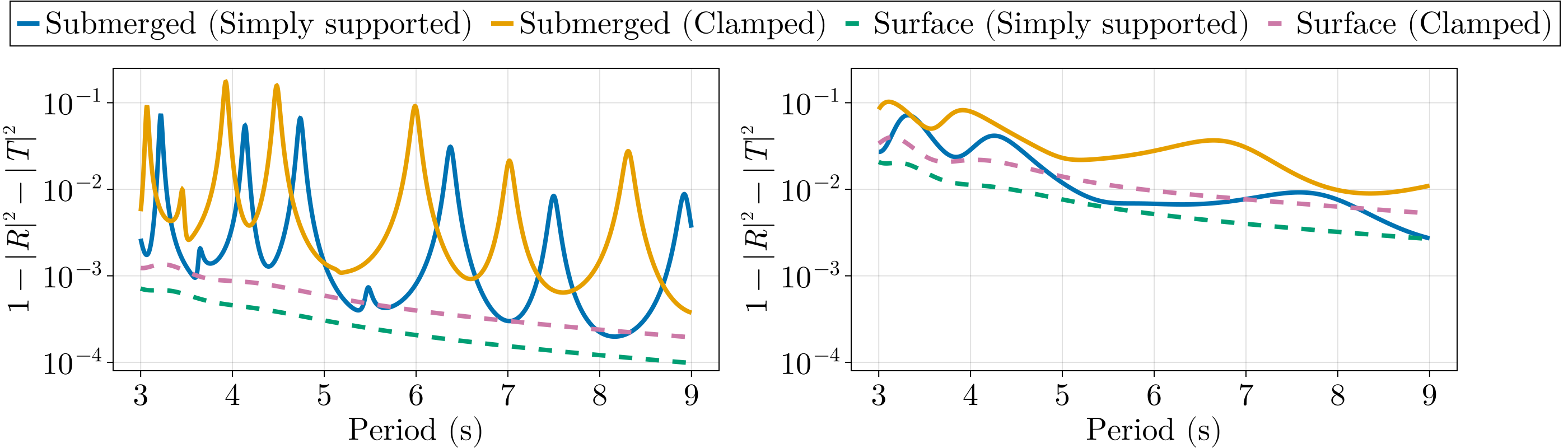}
    \includegraphics[width=\linewidth]{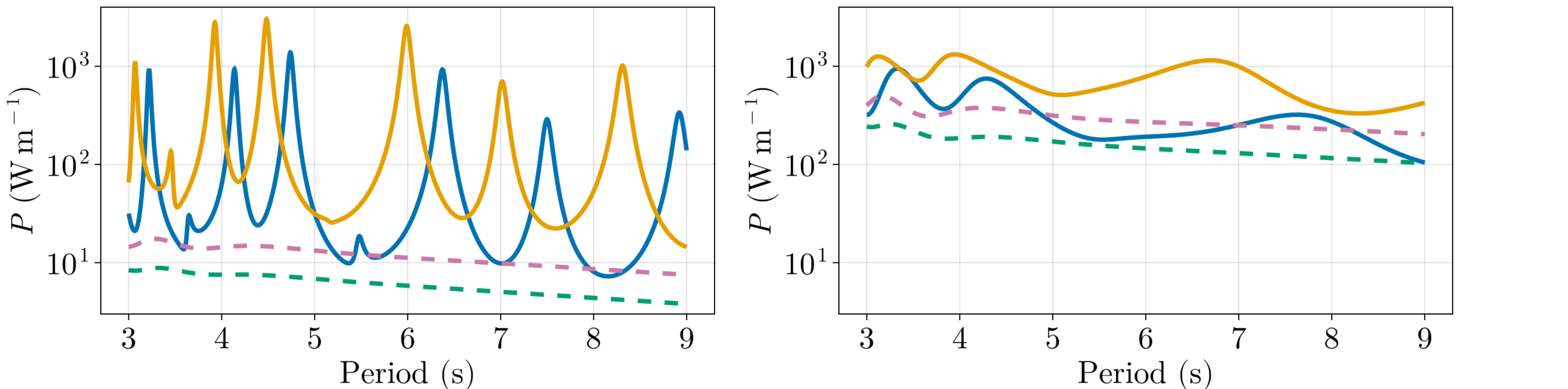}
    \caption{Results for the energy absorption\change{ and power take-off} for periods of 3--9 seconds for the piezoelectric bimorph constructed with (left) PVDF or (right) PZT-5H.}
    \label{fig:pvdf_vs_pzt}
    \footnotesize
    \centering
    \captionof{table}{\change{Energy absorption efficiency of the bimorphs constructed with (a) PVDF and (b) PZT-5H corresponding to the results in Figure~\ref{fig:pvdf_vs_pzt}. The peak period is taken to be $T_p=$ 5 $\mathrm{s}$.}}
    \begin{subtable}[c]{0.45\textwidth}
    \centering
    \change{\begin{tabular}{c|c|c}
        \toprule
        Problem & Boundary condition & Efficiency\\
        \midrule
        Submerged & Simply supported & 0.65\%\\
        Submerged & Clamped & 1.64\%\\
        Surface & Simply supported & 0.039\%\\
        Surface & Clamped & 0.075\%\\
        \bottomrule
    \end{tabular}}
    \caption{\change{PVDF}}
    \label{tab: 2a}
    \end{subtable}
    \hspace{0.4cm}
    \begin{subtable}[c]{0.45\textwidth}
    \centering
    \change{\begin{tabular}{c|c|c}
    \toprule
    Problem & Boundary condition & Efficiency\\
    \midrule
    Submerged & Simply supported & 2.52\%\\
    Submerged & Clamped & 4.68\%\\
    Surface & Simply supported & 1.01\%\\
    Surface & Clamped & 1.85\%\\
    \bottomrule
    \end{tabular}}
    \caption{\change{PZT-5H}}
    \label{tab: 2b}
    \end{subtable}
    \label{tab:pvdf_vs_pzt5a}
\end{figure}
\noindent We now compare PVDF and PZT-5H piezoelectric bimorphs when they are submerged or on the fluid surface for a range of incident wave frequencies and with either clamped or simply supported boundary conditions. Figure~\ref{fig:pvdf_vs_pzt} shows the results for the energy absorption\change{ and power take-off} over wave periods of 3--9 seconds for the parameter values described above. \change{Table~\ref{tab:pvdf_vs_pzt5a} shows the energy absorption efficiency computed for a peak period of $T_p=$ 5 $\mathrm{s}$.} These results show that submerged PWECs are able to absorb much more energy than equivalent PWECs placed on the ocean surface. \change{This is likely caused by the lack of hydrostatic restoring force for the submerged plate, which means that the wave length of the oscillation of the plate is decreased compared to that of the incident wave; hence the second derivative of the out-of-plane displacement is greater than if the plate was on the surface and following the waves. The shorter wave length of the plate's oscillation can be seen in the displacement results in Figure~3.}


Figure~\ref{fig:pvdf_vs_pzt} \change{and Table~\ref{tab:pvdf_vs_pzt5a}} also shows that clamped boundary conditions slightly increase the energy absorbed\change{ and efficiency}, presumably due to the large strains that appear near the end points of the plate, although the effect is not large. Finally, as noted in other works (see \citet{Review_Li_etal2024}), PZT-5H is able to absorb more energy than PVDF because it has larger piezoelectric coupling coefficients. This comes with the trade-off that PZT is more brittle because it is a piezoceramic. It remains to be seen whether this material can be used for practical wave energy applications.

\subsection{Submerged plate depth}
\noindent Having established that submerged PWECs generate more power than surface plates, we now consider varying the depth of submergence $h$. In these experiments, we consider simply supported bimorphs constructed with PVDF or PZT-5H. We use the same parameters as in Sections~\ref{sec: initial results} and \ref{sec: surf vs sub} above, except for the submergence depth. It should be noted that to ensure convergence for small depth ($h<0.5$), $\Delta x$ should be reduced so that it is smaller than the depth. We take $\Delta x=h/10$ for these cases.

\begin{figure}[!t]
    \centering
    \includegraphics[width=\linewidth]{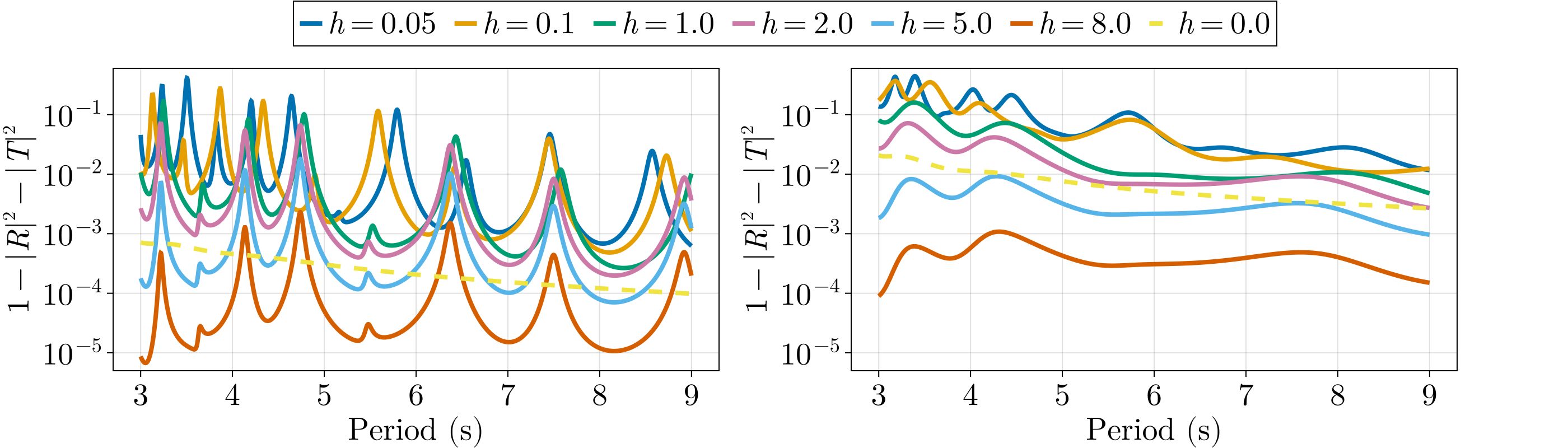}
    \caption{Numerical results for the energy absorption for a simply supported bimorph constructed with (left) PVDF or (right) PZT-5H at various submergence depths. We show the result for the surface case with the dashed line. All values in the legend have units m.}
    \label{fig:depth}
    \captionof{table}{\change{Energy absorption efficiency for simply supported bimorphs constructed of (a) PVDF and (b) PZT-5H at various submergence depths, corresponding to the results in Figure~\ref{fig:depth}. The result for the surface case is given by $h=0$. The peak period is taken to be $T_p=$ 5 $\mathrm{s}$.}}
    \footnotesize
    \centering
    \begin{subtable}[c]{1\textwidth}
    \centering
    \caption{\change{PVDF}}
    \label{tab: 3a}
    \change{\begin{tabular}{c|*{7}{>{\centering\arraybackslash}p{1.1cm}}}
\toprule
Submergence depth ($h$) & 0.00 & 0.05 & 0.10 & 1.00 & 2.00 & 5.00 & 8.00 \\
\midrule
Efficiency & 0.039\% & 2.91\% & 2.18\% & 1.14\% & 0.651\% & 0.142\% & 0.017\% \\
\bottomrule
\end{tabular}}
    \end{subtable}
    \begin{subtable}[c]{1\textwidth}
    \centering
    \caption{\change{PZT-5H}}
    \label{tab: 3b}
    \change{\begin{tabular}{c|*{7}{>{\centering\arraybackslash}p{1.1cm}}}
\toprule
Submergence depth ($h$) & 0.00 & 0.05 & 0.10 & 1.00 & 2.00 & 5.00 & 8.00 \\
\midrule
Efficiency & 1.01\% & 12.7\% & 11.1\% & 5.07\% & 2.52\% & 0.477\% & 0.053\% \\
\bottomrule
\end{tabular}}
    \end{subtable}
    \label{tab:depth}
\end{figure}

Figure~\ref{fig:depth} \change{and Table~\ref{tab:depth}} shows that as the depth of submergence is decreased, the energy absorption \change{and efficiency} increase\change{, respectively}. For large depths, this is expected because the energy in the wave decreases with depth. For very small depths (e.g., $h=0.05\,\mathrm{m}$), the energy absorption is several orders of magnitude greater than the surface plate results (indicated by the dashed lines in Figure~\ref{fig:depth})\change{, however these appear in a narrow frequency band}. This increase in energy absorption has also been shown by other authors \citep[e.g.,][]{Zheng_etal2021,Vipin_Koley_ANN2023,Shoele2023}. Mathematically, it is not possible to take this limit using linear theory. This is demonstrated by the occurrence of an increasing number of resonant peaks as the depth decreases.  In practice, once the upper fluid layer becomes similar in size to the wave amplitude, the linear submergence theory breaks down. \change{In particular, we expect that for $h<1$ non-linear effects will start to dominate and the results in Figure~\ref{fig:depth} for these depths may not be physically meaningful. However, only a non-linear study could quantify this difference.} Experiments have shown that the thin layer of water becomes better modelled by the shallow water equations and that the global motion is well approximated by the floating plate equations (since the thin layer of fluid can be neglected) \cite{meylan2015experimental,skene2015modelling}. This phenomenon should be investigated in future work.

\subsection{Bimorph surface conductance}
\begin{figure}[!t]
    \centering
    \includegraphics[width=\textwidth]{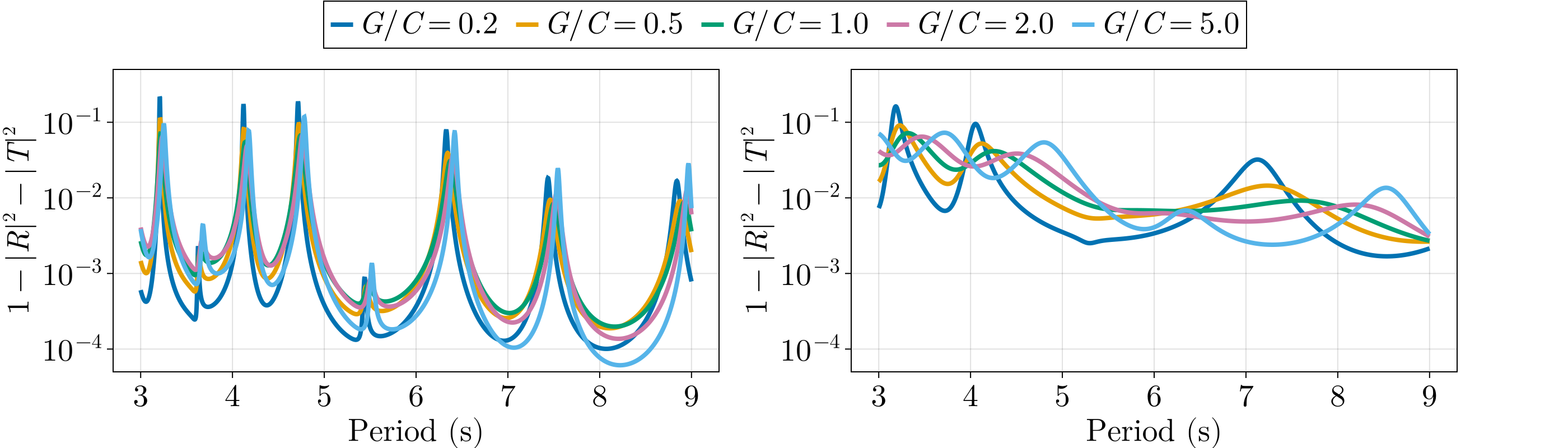}
    \caption{Numerical results for the energy absorption of a simply supported bimorph with (left) PVDF or (right) PZT-5H for varied surface conductance $G$. All values in the legend have units s$^{-1}$.}
    \label{fig:surface conduct}
    \captionof{table}{\change{Energy absorption efficiency for simply supported bimorphs constructed of (a) PVDF and (b) PZT-5H for varied surface conductance, corresponding to the results in Figure~\ref{fig:surface conduct}. The peak period is taken to be $T_p=$ 5 $\mathrm{s}$.}}
    \footnotesize
    \centering
    \begin{subtable}[c]{1\textwidth}
    \centering
    \caption{\change{PVDF}}
    \label{tab: 4a}
    \change{\begin{tabular}{c|*{5}{>{\centering\arraybackslash}p{1.1cm}}}
\toprule
$G/C$ & 0.2 & 0.5 & 1.0 & 2.0 & 5.0 \\
\midrule
Efficiency & 0.599\% & 0.638\% & 0.651\% & 0.653\% & 0.635\% \\
\bottomrule
\end{tabular}}
    \end{subtable}
    \begin{subtable}[c]{1\textwidth}
    \centering
    \caption{\change{PZT-5H}}
    \label{tab: 4b}
    \change{\begin{tabular}{c|*{5}{>{\centering\arraybackslash}p{1.1cm}}}
    \toprule
    $G/C$ & 0.2 & 0.5 & 1.0 & 2.0 & 5.0 \\
    \midrule
    Efficiency & 1.84\% & 2.20\% & 2.52\% & 2.77\% & 3.05\% \\
    \bottomrule
    \end{tabular}}
    \end{subtable}
    \label{tab:surface conduct}
\end{figure}

\noindent Next, we consider the effect on energy absorption of varying the surface conductance $G$. We consider a simply supported bimorph constructed with PVDF or PZT-5H and submerged at a depth of $2\,\mathrm{m}$. Note that in the limits $G\rightarrow0$ and $G\rightarrow\infty$, the energy absorption is zero because the imaginary part of $D$ in Equation~\eqref{eqn: D param} vanishes. Physically, these limits are equivalent to an open circuit or short circuit, respectively. Figure~\ref{fig:surface conduct} \change{and Table~\ref{tab:surface conduct}} shows the energy absorption \change{and efficiency} for various surface conductivities. The results demonstrate that taking $G/C=1$\,s$^{-1}$ (green curve in Fig.~\ref{fig:surface conduct}) gives reasonable energy absorption, however the surface conductance could be tuned to increase the energy absorption for waves of certain periods. \change{For example, for incoming waves of peak period $T_p=$5 $\mathrm{s}$, Table~\ref{tab:surface conduct} shows that $G/C=2$\,s$^{-2}$ and $G/C=5$\,s$^{-2}$ give the best energy absorption efficiency for PVDF and PZT-5H, respectively.}

\subsection{Poling angle}
\begin{figure}[!t]
    \centering
    \includegraphics[width=\textwidth]{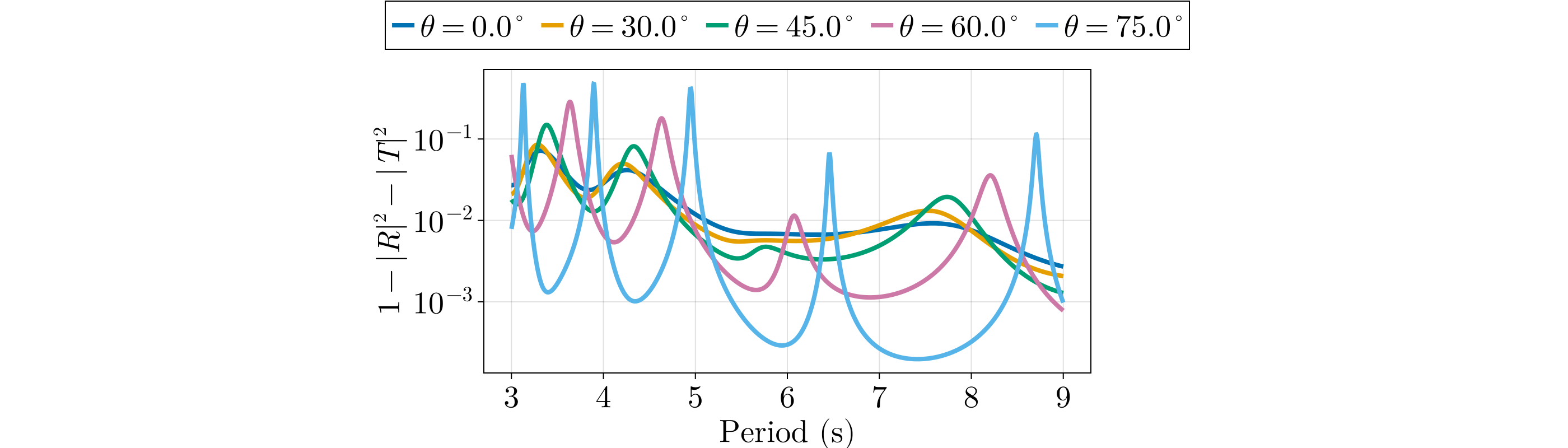}
    \caption{Numerical results for the energy absorption of a simply supported bimorph made of PZT-5H with varied poling angle $\theta$.}
    \label{fig:poling angle}
    \centering
    \captionof{table}{\change{Energy absorption efficiency of a simply supported bimorph made of PZT-5H with varied poling angle $\theta$, corresponding to the results in Figure~\ref{fig:poling angle}. The peak period is taken to be $T_p=$ 5 $\mathrm{s}$.}}
    \label{tab:poling angle}
    \footnotesize
    \begin{subtable}[c]{\textwidth}
    \centering
    \change{\begin{tabular}{c|*{5}{>{\centering\arraybackslash}p{1.1cm}}}
\toprule
Poling angle $\theta$ & $0^\circ$ & $30^\circ$ & $45^\circ$ & $60^\circ$ & $75^\circ$ \\
\midrule
Efficiency & 2.52\% & 2.42\% & 2.77\% & 3.14\% & 2.52\% \\
\bottomrule
\end{tabular}}
    \end{subtable}
\end{figure}

\noindent In this section, we vary the poling angle of the piezoelectric layers in the bimorph, and instead of the two layers being oppositely poled with poling in the vertical direction, the poling direction is rotated clockwise by angle $\theta$. We consider a simply supported bimorph constructed with PZT-5H submerged at a depth of $2\,\mathrm{m}$. We do not consider PVDF in this case because the change in poling angle requires the full piezoelectric material tensors. For these experiments we fix the ratio $G/C=1\,\mathrm{s}^{-1}$. It should be noted that a poling angle of $90^\circ$ corresponds to poling in the $x$-direction (i.e., parallel to the plate). In this case, the imaginary part of $D$ again vanishes and thus no energy is absorbed. Figure~\ref{fig:poling angle} \change{and Table~\ref{tab:poling angle}} shows the numerical results for the energy absorption \change{and efficiency} for various poling angles. The results show that while peaks appear in energy absorption spectrum, the energy absorbed typically decreases as the rotation of the poling direction increases towards $90^\circ$. \change{Table~\ref{tab:poling angle} demonstrates that for incoming waves of peak period $T_p=$ 5 $\mathrm{s}$, adjusting the poling angle to $60^\circ$ yields nearly a 25\% improvement in the energy absorption efficiency. This suggests that varying the poling angle may provide practical benefit to energy absorption. More generally, we expect that varying the piezoelectric material orientation could improve energy absorption for three-dimensional anisotropic plates. Adjusting the poling angle and material orientation should continue to be investigated in future studies.}

\section{Conclusions}\label{conclusions_sec}
\noindent In this paper we have numerically investigated wave energy absorption by piezoelectric bimorphs when submerged or placed on the surface of the ocean. In the first part of the paper we carefully derived the equations of motion in dimensional form starting from the constitutive relations for piezoelectricity. This allowed us to consider realistic piezoelectric material properties \change{and extend the model to } arbitrary poling angle for the two oppositely-poled piezoelectric layers. Our derivation verifies the non-dimensional model developed by \citet{renzi2016hydroelectromechanical} and gives additional details. Although we do not describe it here, the derivation can also be extended to three-dimensional problems.

To solve the coupled system in Equations~\eqref{eqn: coupled system} for the submerged plate we implemented a modal expansion method. The component diffraction and radiation problems were reduced to hypersingular integral equations and solved numerically using a constant panel method. A key step in this approach is the numerical integration of the derivative of the Green's function, which is achieved using residue calculus and numerical integration over the resulting deformed contour. The method is general and can be extended to problems involving plates submerged vertically or at an angle, curved plates, and three-dimensional problems. We note that the models are limited by the assumption that the plate thickness can be neglected and that the wave amplitude and plate motion \change{are} small.

We considered several numerical experiments using the above approach. In the first numerical study, we computed the energy absorption \change{and efficiency for a peak period of 5 $\mathrm{s}$} for piezoelectric bimorphs, with the piezoelectric layer consisting of PVDF or PZT-5H. We compared energy absorption \change{and efficiency} when the plate is submerged or on the surface and showed that much greater energy absorption is obtained using submerged plates, particularly for bimorphs constructed with PZT-5H. We considered clamped and simply supported boundary conditions. We found that these give qualitatively similar results with a slight increase in energy absorption for clamped plates because of the large strains that appear near the end points of the plate. 

In the remaining numerical experiments, we considered varying the plate submergence depth, surface conductance, and poling angle. The results for varied plate submergence depth showed that for small depths, the energy absorption is several orders of magnitude larger than the surface plate results. We know that as the plate approaches the surface there is no stable limit, and the response becomes dominated by resonant waves on the top of the plate. Moreover, no matter how close to the surface, the hydrodynamic restoring force is missing from the equations of motion of the submerged plate.  There is significant scope for further investigation of this phenomena and it is not clear whether the submerged plate works better in absorbing energy for all parameter values. The results for varying the surface conductance show that this is a tunable parameter for increasing energy absorption \change{and efficiency} at certain wave periods. \change{Finally, we found that varying the poling angle can provide practical benefit by increasing the energy absorption efficiency, with a near 25\% improvement for waves with a peak period of 5 $\mathrm{s}$.}

\section*{Data availability}
\noindent The source code and data for this work is available at \url{https://github.com/zjwegert/SemiAnalyticWECs.jl}.

\section*{Acknowledgment}

\noindent This research supported by the Australian Government through the Australian Research Council (ARC), Discovery Grant DP240102104.

\appendix

\section{Expressions for material constants}\label{appendix1}
\noindent In the following, we give the expressions for $\hat{C}^E_{11}(\theta)$, $\hat{e}_{31}(\theta)$ and $\hat{\kappa}_{33}^\sigma(\theta)$ in terms of components of the piezoelectric material tensors $\bar{S}^E_{pq}$, $\bar{d}_{ip}$, and $\bar{\kappa}^\sigma_{ij}$ in the original reference frame. To shorten the expressions, we denote $\mathrm{s}=\sin\theta$ and $\mathrm{c}=\cos\theta$. These expressions are given by

\begin{equation}
\begin{aligned}
\hat{C}^E_{11}(\theta)&=8 \bar{S}^E_{22}/{\large{[}}\bar{S}^E_{22} \bar{S}^E_{55}-s^4 (8 [\bar{S}^E_{23}]^2-8 \bar{S}^E_{22} \bar{S}^E_{33}+\bar{S}^E_{22} \bar{S}^E_{55})+2 c^2 s^2 (8 \bar{S}^E_{13} \bar{S}^E_{22}-8 \bar{S}^E_{12} \bar{S}^E_{23}+3 \bar{S}^E_{22} \bar{S}^E_{55})\\&\quad\quad-c^4 (8 [\bar{S}^E_{12}]^2+\bar{S}^E_{22} (-8 \bar{S}^E_{11}+\bar{S}^E_{55})){\large{]}},
\end{aligned}
\end{equation}
\begin{equation}
\begin{aligned}
\hat{e}_{31}(\theta)&={\large{[}}8 (c^3 (\bar{d}_{32} \bar{S}^E_{12}-\bar{d}_{31} \bar{S}^E_{22})+c s^2 (\bar{d}_{15} \bar{S}^E_{22}-\bar{d}_{33} \bar{S}^E_{22}+\bar{d}_{32} \bar{S}^E_{23})){\large{]}}/{\large{[}}-\bar{S}^E_{22} \bar{S}^E_{55}\\&\quad\quad+s^4 (8 [\bar{S}^E_{23}]^2-8 \bar{S}^E_{22} \bar{S}^E_{33}+\bar{S}^E_{22} \bar{S}^E_{55})-2 c^2 s^2 (8 \bar{S}^E_{13} \bar{S}^E_{22}-8 \bar{S}^E_{12} \bar{S}^E_{23}+3 \bar{S}^E_{22} \bar{S}^E_{55})\\&\quad\quad+c^4 (8 [\bar{S}^E_{12}]^2+\bar{S}^E_{22} (-8 \bar{S}^E_{11}+\bar{S}^E_{55})){\large{]}},
\end{aligned}
\end{equation}
\begin{equation}
\begin{aligned}
\hat{\kappa}_{33}^\sigma(\theta)&=\frac{1}{\bar{S}^E_{22}}{\large{[}}-c^2 [\bar{d}_{32}]^2+c^2 \bar{\kappa}^\sigma_{33} \bar{S}^E_{22}+\bar{\kappa}^\sigma_{11} s^2 \bar{S}^E_{22}+\{8 (c^3 (\bar{d}_{32} \bar{S}^E_{12}-\bar{d}_{31} \bar{S}^E_{22})\\&\quad\quad+c s^2 (\bar{d}_{15} \bar{S}^E_{22}-\bar{d}_{33} \bar{S}^E_{22}+\bar{d}_{32} \bar{S}^E_{23}))^2\}/\{-\bar{S}^E_{22} \bar{S}^E_{55}+s^4 (8 [\bar{S}^E_{23}]^2-8 \bar{S}^E_{22} \bar{S}^E_{33}+\bar{S}^E_{22} \bar{S}^E_{55})\\&\quad\quad-2 c^2 s^2 (8 \bar{S}^E_{13} \bar{S}^E_{22}-8 \bar{S}^E_{12} \bar{S}^E_{23}+3 \bar{S}^E_{22} \bar{S}^E_{55})+c^4 (8 [\bar{S}^E_{12}]^2+\bar{S}^E_{22} (-8 \bar{S}^E_{11}+\bar{S}^E_{55}))\}{\large{]}}.
\end{aligned}
\end{equation}
For PZT-5H, the piezoelectric material tensors in the reference frame with poling in the $z$-direction are \citep{YangIntroPZ}
\begin{subequations}
\begin{align}
    \bar{\boldsymbol{S}}^E_{\mathrm{PZT5H}} &= \begin{pmatrix}
         16.7  & -4.84 & -8.5 & 0.0 &  0.0 &  0.0\\
         -4.84 & 16.7  & -8.5 & 0.0 &  0.0 &  0.0\\
         -8.5  & -8.5  & 20.8 & 0.0 &  0.0 &  0.0\\
          0.0  &  0.0  &  0.0 & 43.5 &  0.0 &  0.0\\
          0.0  &  0.0  &  0.0 & 0.0 & 43.5 &  0.0\\
          0.0  &  0.0  &  0.0 & 0.0 &  0.0 & 42.9
    \end{pmatrix}\times10^{-12}~~\mathrm{(m^2N^{-1})},\\
    \bar{\boldsymbol{d}}_{\mathrm{PZT5H}} &= \begin{pmatrix}
          0.0  &  0.0  & 0.0  & 0.0  & 7.39 & 0.0\\
          0.0  &  0.0  & 0.0  & 7.39 & 0.0  & 0.0\\
         -2.75 & -2.75 & 5.94 & 0.0  & 0.0  & 0.0
    \end{pmatrix}\times10^{-10}~~\mathrm{(mV^{-1})},\\
    \bar{\boldsymbol{\kappa}}^\sigma_{\mathrm{PZT5H}} &=\begin{pmatrix}
        2.76&0&0\\
        0&2.76&0\\
        0&0&3.04
    \end{pmatrix}\times10^{-8}~~\mathrm{(Fm^{-1})}.
\end{align}
\end{subequations}

\section{Calculation of the Green's function}\label{greens_fn_appendix}
\noindent The Green's function in Equation~\eqref{eqn:Green's function} requires the evaluation of the integral
\begin{align}\label{appendix_integral}
I(X) = -2\Uint_0^\infty
{\cos(\mu X)}F(\mu)\upd\mu,
\end{align}
with
\begin{equation}
    F(\mu)=\frac{1}{\cosh \mu H}
\left[\frac{\cosh\mu(z+H)\cosh\mu(z^\prime+H)}{\mu\sinh\mu H-\alpha\cosh \mu H}+\frac{\e^{-\mu H}}{\mu}\sinh\mu z\sinh \mu z^\prime\right]. 
\end{equation}
This is not suitable for numerical calculation due to the pole at $\mu=k$, where we recall $k$ satisfies $k\sinh k H-\alpha\cosh k H=0$. We first write Equation \eqref{appendix_integral} as $I(X)=I_+(X)+I_-(X)$ where
\begin{equation}
I_\pm(X) = -\Uint_0^\infty
\mathrm{e}^{\pm\mathrm{i} \mu X}F(\mu)\upd\mu.
\end{equation}
\begin{figure}[!t]
    \centering
    \def\svgwidth{0.4\textwidth}
    \input{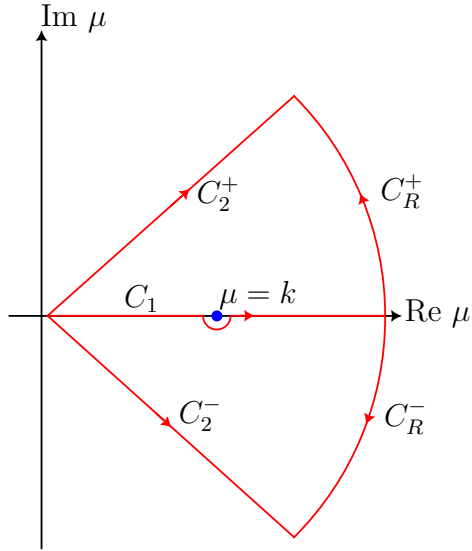}
    \caption{Deformed integration contour}
    \label{fig:contour}
\end{figure}
We assume that $X>0$, noting that the case $X<0$ can be recovered from the relationship $I(-X)=I(X)$. The assumption $X>0$ allows us to deform the contour of integration for $I_\pm(X)$ to be over $C^\pm=C_R^{\pm}\cup C_2^{\pm}$ as $R\rightarrow\infty$. We parameterise $C_2^{\pm}$ as $\gamma(R)=Re^{\pm\pi\upi/4}$ for $R\in[0,\infty)$. Figure~\ref{fig:contour} shows a visualisation of this. When closing the contour for $I_+(X)$ in the upper-half plane, we pick up a contribution from the pole at $\mu=k$. It can be shown that as $R\rightarrow\infty$, the contributions from $C_R^{\pm}$ to $I_\pm(X)$ vanish. As $R\rightarrow\infty$, we are left with
\begin{equation}
    I_+(X)=-\int_{C_2^+}
\mathrm{e}^{\mathrm{i} \mu X}F(\mu)\upd\mu-2\pi\upi\mathop{\mathrm{Res}}_{\mu = k}\left\{e^{\upi\mu X}F(\mu)\right\},
\end{equation}
and 
\begin{equation}
    I_-(X)=-\int_{C_2^-}
\mathrm{e}^{-\mathrm{i} \mu X}F(\mu)\upd\mu,
\end{equation}
where the residue is given by \citep{Linton2001}
\begin{equation}
    2\pi\upi\mathop{\mathrm{Res}}_{\mu = k}\left\{e^{\upi\mu X}F(\mu)\right\}=\frac{\pi\upi}{kH(1+\sinh(2kH)/2kH)}\cosh(k(z+H))\cosh(k(z^\prime+H))e^{\upi k X}.
\end{equation}
Finally, we observe that the integrals appearing in the expressions for $I_+(X)$ and $I_-(X)$ are complex conjugates, so we only require one to be calculated and this can be done with numerical quadrature. This completes the calculation of the Green's function.

\section{Validation for the case of a perfectly elastic plate}
\label{comparison}

\begin{figure}[!t]
    \centering
    \includegraphics[width=0.8\textwidth]{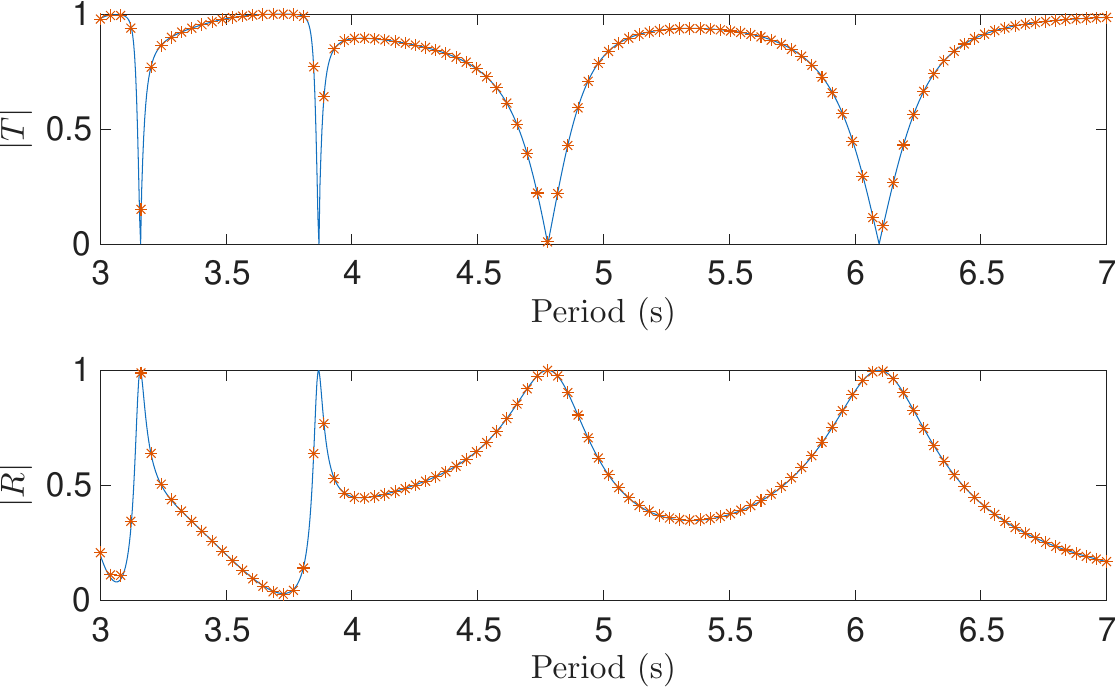}
    \caption{Validation of the reflection and transmission coefficients resulting from the hypersingular boundary integral equation method (solid line) presented in this paper and eigenfunction matching (stars) using the method in \cite{meylan2009water}}.
    \label{fig:validation}
\end{figure}

\noindent We present here a validation of the hypersingular boundary integral equation method for the case of a submerged perfectly elastic plate. This provides confidence that the code is correct and also provides a benchmark solution that may help with the development of other codes in the future.  The presented method is compared with the solution by eigenfunction matching developed by \citet{meylan2009water}. For this calculation we take the water depth as $H=2.5$\,m, the depth of submergence as $h=0.25$\,m, the water density as $1000$\,kg\,m$^{-3}$, the flexural rigidity as $D=10^4$\,N\,m$^{2}$, and the mass moment of inertia to be $I_b=10$\,kg\,m$^{-2}$. Figure~\ref{fig:validation} shows the results of this comparison, where we plot the absolute value of the reflection and transmission coefficients for wave periods between 3 and 7 seconds. 

\section{\changerevtwo{Validation of energy conservation}}\label{sec:val coe}
\begin{figure}[!t]
    \centering
    \includegraphics[width=\linewidth]{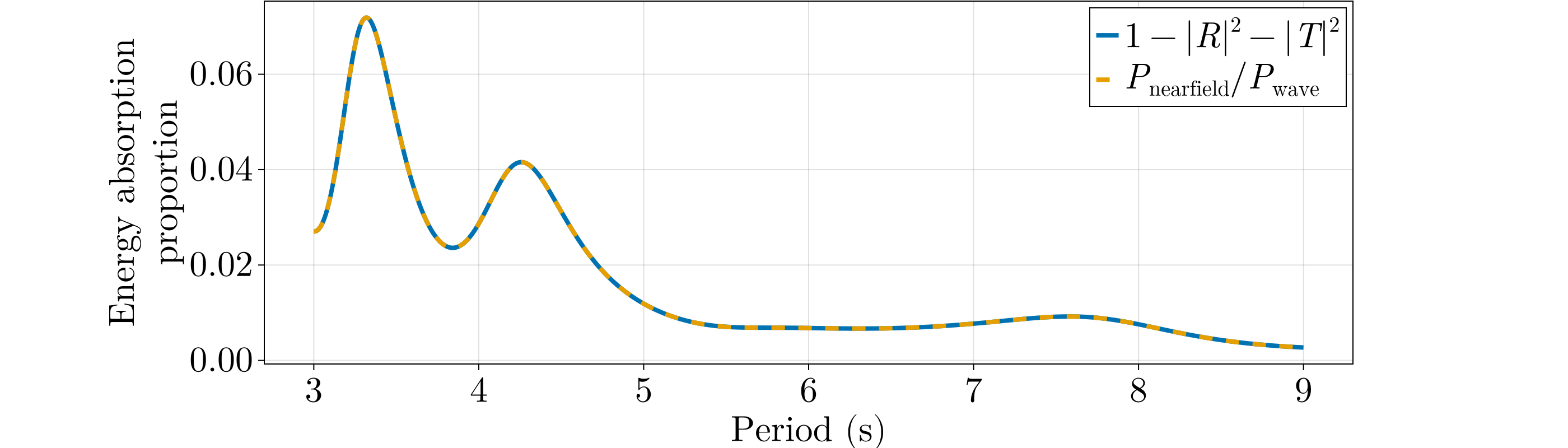}
    \caption{\changerevtwo{Proportion of energy absorbed by a simply supported bimorph made of PZT-5H computed from the transmission and reflection coefficients (Eqn.~\ref{eqn: energy absorb}) or the nearfield power (Eqn.~\ref{eqn: nearfield power}) divided by the wave power. The power in the wave is given by $P_{\mathrm{wave}}$.}}
    \label{fig:validation-coe}
\end{figure}

\noindent \changerevtwo{We present here a validation of energy conservation for the case of a simply supported bimorph made of PZT-5H. The parameter values used are as in Section~\ref{sec: initial results}. Figure~\ref{fig:validation-coe} shows the proportion of energy absorbed calculated using either: i) the transmission and reflection coefficients (Eqn.~\ref{eqn: energy absorb}); or ii) the nearfield power (Eqn.~\ref{eqn: nearfield power}) divided by the wave power, $P_{\mathrm{wave}}=\frac{1}{2}\rho_wgA^2C_g$.}

\biboptions{sort&compress}
\bibliographystyle{elsarticle-num-names} 
\bibliography{refs}





\end{document}